\documentclass{emulateapj}

\newcommand{\lya}{Ly$\alpha$}

\begin{document}

\title{PHYSICAL PROPERTIES OF SPECTROSCOPICALLY-CONFIRMED GALAXIES AT $z\ge6$. 
II. MORPHOLOGY OF THE REST-FRAME UV CONTINUUM AND Ly-$\alpha$ 
EMISSION\footnotemark[$\ast$]}
\footnotetext[$\ast$]{
Based in part on observations made with the NASA/ESA Hubble Space Telescope,
obtained from the data archive at the Space Telescope Science Institute, which
is operated by the Association of Universities for Research in Astronomy, Inc.
under NASA contract NAS 5-26555.
Based in part on observations made with the Spitzer Space Telescope, which is
operated by the Jet Propulsion Laboratory, California Institute of Technology
under a contract with NASA.
Based in part on data collected at Subaru Telescope and obtained from the
SMOKA, which is operated by the Astronomy Data Center, National Astronomical
Observatory of Japan.}

\author{Linhua Jiang\altaffilmark{1,2,9}, Eiichi Egami\altaffilmark{2},
Xiaohui Fan\altaffilmark{2}, Rogier A. Windhorst\altaffilmark{1},
Seth H. Cohen\altaffilmark{1}, Romeel Dav\'{e}\altaffilmark{2,3},
Kristian Finlator\altaffilmark{4,10}, Nobunari Kashikawa\altaffilmark{5},
Matthew Mechtley\altaffilmark{1}, Masami Ouchi\altaffilmark{6,7},
and Kazuhiro Shimasaku\altaffilmark{8}}
\altaffiltext{1}{School of Earth and Space Exploration, Arizona State
   University, Tempe, AZ 85287-1504, USA; linhua.jiang@asu.edu}
\altaffiltext{2}{Steward Observatory, University of Arizona,
   933 North Cherry Avenue, Tucson, AZ 85721, USA}
\altaffiltext{3}{Physics Department, University of the Western Cape, 7535
   Bellville, Cape Town, South Africa}
\altaffiltext{4}{Dark Cosmology Centre, Niels Bohr Institute, University of 
Copenhagen, DK-2100 Copenhagen $\O$, Denmark}
\altaffiltext{5}{Optical and Infrared Astronomy Division, National
   Astronomical Observatory, Mitaka, Tokyo 181-8588, Japan}
\altaffiltext{6}{Institute for Cosmic Ray Research, The University of Tokyo,
   5-1-5 Kashiwanoha, Kashiwa, Chiba 277-8582, Japan}
\altaffiltext{7}{Kavli Institute for the Physics and Mathematics of the
   Universe, The University of Tokyo, 5-1-5 Kashiwanoha, Kashiwa, Chiba
   277-8583, Japan}
\altaffiltext{8}{Department of Astronomy, University of Tokyo, Hongo, Tokyo
   113-0033, Japan}
\altaffiltext{9}{Hubble Fellow.}
\altaffiltext{10}{DARK fellow.}

\begin{abstract}

We present a detailed structural and morphological study of a large sample of 
spectroscopically-confirmed galaxies at $z\ge6$, using deep $HST$ near-IR
broad-band images and Subaru optical narrow-band images. The galaxy sample 
consists of 51 \lya\ emitters (LAEs) at $z\simeq5.7$, 6.5, and 7.0, and 16 
Lyman-break galaxies (LBGs) at $5.9\le z\le6.5$. These galaxies exhibit a wide 
range of rest-frame UV continuum morphology in the $HST$ images, from compact 
features to multiple component systems. The fraction of merging/interacting 
galaxies reaches 40\% $\sim$ 50\% at the brightest end of $M_{1500}\le-20.5$ 
mag. The intrinsic half-light radii $r_{\rm hl,in}$, after correction for PSF 
broadening, are roughly between $r_{\rm hl,in}\simeq0\farcs05$ (0.3 kpc) and 
$0\farcs3$ (1.7 kpc) at $M_{1500}\le-19.5$ mag. The median $r_{\rm hl,in}$ 
value is $0\farcs16$ ($\sim$0.9 kpc). This is consistent with the sizes of 
bright LAEs and LBGs at $z\ge6$ in previous studies. In addition, more 
luminous galaxies tend to have larger sizes, exhibiting a weak size-luminosity 
relation $r_{\rm hl,in} \propto L^{0.14}$ at $M_{1500}\le-19.5$ mag. The slope 
of 0.14 is significantly flatter than those in fainter LBG samples. 
We discuss the morphology of $z\ge6$ galaxies with nonparametric methods,
including the $CAS$ system and the Gini and $M_{20}$ parameters, and 
demonstrate their validity through simulations.
We search for extended \lya\ emission halos around LAEs at $z\simeq5.7$ and 
6.5, by stacking a number of narrow-band images. We do not find evidence of 
extended halos predicted by cosmological simulations. Such \lya\ halos, if 
they exist, could be weaker than predicted.
Finally, we investigate any positional misalignment between UV continuum and 
\lya\ emission in LAEs. While the two positions are generally consistent,
several merging galaxies show significant positional differences. This is 
likely caused by a disturbed ISM distribution due to merging activity.

\end{abstract}


\section{INTRODUCTION}

Galaxy structural and morphological studies provide basic apparent information 
about galaxies. Nearby galaxies are generally classified into three broad 
categories: spiral, elliptical, and irregular. The majority of 
luminous nearby galaxies ($z\le0.1$) are spirals and ellipticals 
\citep{abr01}. At higher redshift, galaxies are less well developed, and the 
fraction of irregular galaxies increases steadily 
\citep[e.g.,][]{driver95,driver98}. In the redshift range of $0.1\le z\le1$, 
galaxy morphology and structure have also been well studied
\citep[e.g.,][and references therein]{bri98,lil98,sch99,car00,lef00,van00}.
These galaxies show more disturbed structures than nearby galaxies do in the
rest-frame UV and optical \citep[e.g.,][]{abr01,win02,taylor07,bla09,shi09}. 
The fraction of irregular galaxies increases from less than 10\% at $z\le0.5$ 
to $\sim30$\% at $z\simeq1$ \citep[e.g.,][]{bri98,van00}. In addition, more 
galaxies were identified as merging systems, reflecting the hierarchical 
build-up of galaxies and mass assembly in the cold dark matter (CDM) scenario 
\citep{whi78,col00}. For example, nearly 20\% of $z\simeq1$ galaxies in the 
\citet{lef00} sample are in close pairs.

For galaxies at $z\ge2-3$, morphological classification is challenging, as 
most galaxies appear peculiar. Galaxies are also smaller towards higher 
redshift \citep{fer04}. In addition, galaxies appear much fainter due to the 
cosmological $(1+z)^4$ surface brightness dimming. Traditional 
classifications, including Hubble's tuning-fork system, are no longer 
practical at these higher redshifts. 
Therefore, nonparametric methods such as the $CAS$ system 
\citep{con03} and the Gini and $M_{20}$ parameters \citep{lot04} play an 
important role. Most morphological and structural analyses in this redshift 
range were done in the GOODS fields \citep{gia04}, because of the high-quality 
{\it Hubble Space Telescope} ($HST$) data 
\citep[e.g.,][]{low97,rav06,law07,cas10}. In the rest-frame UV, $z\ge3$ 
galaxies are usually compact (from one to several kpc), but many of 
them display extended features or multiple clumps in deep $HST$ images 
\citep[e.g.,][]{gia96,ven05,rav06,pir07,con09,cooke10,gronwall11,law12}. 
For example, the \citet{rav06} sample contains thousands of 
photometrically-selected LBGs at $z\ge2.5$, and about 30\% of them have 
multiple cores. In \citet{law12}, a sample of spectroscopically-confirmed 
galaxies in a similar redshift range also showed a high fraction of 
interacting systems.

In the highest-redshift range $z\ge6$, morphological studies become even
more difficult. 
Galaxies appear very faint, and their rest-frame UV light moves to the near-IR 
wavelength range, where telescope resolution is poorer. A typical galaxy 
occupies only a few pixels even in $HST$ near-IR images, so size is usually 
the only physical parameter that can be reliably measured in the literature. 
Studies based on photometrically-selected galaxies have shown that $z\ge6$ 
galaxies are 
generally very compact, and most of them are just barely spatially resolved. 
For example, \citet{oes10} reported the sizes of 16 LBGs at $z\ge7$ in the 
Hubble Ultra-Deep Field (HUDF). They found that only two in their sample show 
extended features, and the rest are very compact ($\le1$ kpc). This sample is 
very faint. Observations of a handful brighter galaxies with spectroscopic 
redshifts also suggest compact morphology, with a typical size of $\le1$ kpc
\citep[e.g.,][]{sta04,dow07,cow11}. Note that galaxy size is correlated with 
physical properties, such as mass and luminosity at low redshift. Such 
relations may still exist in high-redshift galaxies, but could have evolved 
with time \citep[e.g.,][]{gra12,mos12}.

In this paper, we will carry out a structural and morphological study of a 
sample of 67 galaxies at $z\ge6$. The sample is the largest collection of 
spectroscopically-confirmed galaxies in this redshift range, including 51 
\lya\ emitters (LAEs) and 16 Lyman-break galaxies (LBGs). This paper is the 
second in a series presenting the physical properties of these galaxies. 
In the first paper of the series \citep[hereafter Paper I]{jia13}, we 
presented deep Subaru optical and $HST$ near-IR data. We also derived various 
rest-frame UV continuum and \lya\ emission properties, including UV-continuum 
slope $\beta$, the \lya\ rest-frame equivalent width (EW), and star formation 
rates (SFRs). These galaxies have steep UV continuum slopes roughly between 
$\beta\simeq-1.5$ and $-3.5$, with a mean value of $\beta\simeq-2.3$. They have 
a range of \lya\ EW from $\sim$10 to $\sim$200 
\AA. Their SFRs are moderate from a few to a few tens solar masses per year.
In this paper, we will study the structure and morphology of their rest-frame 
UV-continuum emission based on our $HST$ images, and of their \lya\ emission 
based on our ground-based narrow-band images.

The layout of the paper is as follows. In Section 2, we briefly review our 
galaxy sample and the optical and near-IR data that will be used for the 
paper. We measure the structure and morphology of UV continuum emission in 
Section 3, and of the \lya\ emission in Section 4. We then discuss our results
and summarize the paper in Section 5.
Throughout the paper we use a $\Lambda$-dominated flat cosmology with $H_0=70$
km s$^{-1}$ Mpc$^{-1}$, $\Omega_{m}=0.3$, and $\Omega_{\Lambda}=0.7$ 
\citep{kom11}. All magnitudes are on the AB system \citep{oke83}.

\section{GALAXY SAMPLE AND DATA}

In Section 2 of Paper I, we provided a detailed description of our galaxy 
sample and the multi-wavelength data that we studied. Here we summarize
the information below. There are a total of 67 spectroscopically-confirmed 
galaxies in our sample: 62 are from the Subaru Deep Field 
\citep[SDF;][]{kas04}, and the remaining 5 are from the Subaru XMM-Newton Deep 
Survey field \citep[SXDS;][]{fur08}. They represent the most luminous galaxies 
in terms of \lya\ luminosity (for LAEs) or UV continuum luminosity (for LBGs) 
in this redshift range.
The SDF galaxy sample contains 22 LAEs at $z\simeq5.7$ \citep{shi06,kas11},
25 LAEs at $z\simeq6.5$ \citep{tan05,kas06,kas11}, and a LAE at $z=6.96$
\citep{iye06}. The LAEs at $z\simeq5.7$ and 6.5 have a relatively uniform
magnitude limit of 26 mag in the narrow bands NB816 and NB921, and thus make
a well-defined sample. The SDF sample also contains 14 LBGs at
$5.9\le z\le6.5$ \citep{nag04,nag05,nag07,ota08,jia11,tos12}. The LBG 
candidates in these studies were
selected with different criteria, and have a rather inhomogeneous depth.
The SXDS sample consists of five galaxies, including two LBGs at $z\simeq6$
\citep{cur12} and three LAEs at $z\simeq6.5$ \citep{ouc10}.
All these galaxies are listed in Table 1 of Paper I.

The SDF and SXDS were observed with Subaru Suprime-Cam \citep{kas04,fur08}.
They have extremely deep optical images in a series of broad and narrow bands.
Public stacked images are available for the two fields, but the public
data do not include the images taken recently. In Paper I, we produced our own
stacked images in six broad bands ($BVRi'z'y$) and three narrow bands (NB816,
NB921, and NB973) by including all available data in the archive. Our stacked
images have great depth with excellent PSF full width at half-maxima (FWHMs) 
of $0\farcs5-0\farcs7$. The near-IR imaging data for the SDF galaxies are from 
three $HST$ GO programs 11149 (PI: E. Egami), 12329 and 12616 (PI: L. Jiang). 
The $HST$ observations were made with a mix of instruments and depth. 
The majority of the galaxies were observed with WFC3 in the F125W (hereafter 
$J_{125}$) and F160W (hereafter $H$ or $H_{160}$) bands. The typical
integration time was two $HST$ orbits (roughly 5400 s) per band. This
provides a depth of $\sim27.5$ mag ($5\sigma$ detection) in the $J_{125}$ band
and $\sim27.1$ mag in the $H_{160}$ band \citep[see also][]{win11}. The pixel 
size in the final reduced WFC3 images is $0\farcs06$. Several SDF galaxies 
were observed with NICMOS in the F110W (hereafter $J_{110}$) and $H_{160}$ 
bands. The typical integration time was also two $HST$ orbits. The depth in 
the two bands are $\sim26.4$ mag and $\sim26.1$ mag, respectively. The pixel 
size in the final reduced NICMOS images is $0\farcs1$. The five SXDS galaxies 
were covered by the UKIDSS Ultra-Deep Survey (UDS). Their $HST$ WFC3 near-IR 
data were obtained from the Cosmic Assembly Near-infrared Deep Extragalactic 
Legacy Survey \citep[CANDELS;][]{grogin11,koe11}.
The exposure depth of the CANDELS UDS data is 1900 s in the $J_{125}$ band
and 3300 s in the $H_{160}$ band.

The majority of the galaxies in our sample were detected with high 
significance in the near-IR images. Only 15 of them --- among the faintest in 
the optical --- have weak detections ($<5\sigma$) in the $J$ band ($J_{125}$ 
or $J_{110}$). In Table 1 of Paper I, we listed the optical and near-IR 
photometry of the galaxies. In Table 2 of Paper I, we presented basic physical 
properties, including the rest-frame UV continuum luminosity and slope 
$\beta$, the \lya\ luminosity and EW, and SFR, etc. The thumbnail 
images of all the galaxies are provided in Appendix A of Paper I.

\section{UV CONTINUUM MORPHOLOGY}

In this section, we will derive structural and morphological parameters for 
the galaxies in our sample. Although our galaxies represent the most luminous 
galaxies at $z\ge6$, they appear faint and small compared to lower-redshift 
galaxies. The majority of them are point-like sources in the Subaru optical 
images. Even in the $HST$ WFC3 images they usually occupy a very limited 
number of pixels. Hence, the study of these distant galaxies is challenging. 
In previous literature, galaxy size was often the only parameter that could be 
reliably measured for $z\ge6$ galaxies. Classifications for nearby galaxies,
such as the classical Hubble's tuning-fork system, cannot be applied to these 
objects. In this section, we will measure the sizes of our galaxies, and try 
to characterize their morphology using nonparametric methods, such as the 
$CAS$ system \citep{con03} and the Gini and $M_{20}$ parameters \citep{lot04}. 
These methods are primarily used for low-redshift galaxies, though they have 
already been used for galaxies at $z=4\sim6$ \citep[e.g.,][]{pir07,con09}. 
We will also study interacting/merging systems in our sample.

In order to calculate the above parameters, we took advantage of all our $HST$ 
images. For each galaxy in our sample, we combined (the weighted average) 
its $J$- and $H$-band images and made a stacked $HST$ image to improve the
signal-to-noise (S/N) ratio. By doing this we assume that the effect of the 
morphological k-correction --- the dependence of galaxy structure on 
wavelength --- is negligible in the wavelength range considered.
This is because the $J$ and $H$ bands cover a similar rest-frame UV wavelength 
range ($\sim$1780 \AA\ vs. $\sim$2200 \AA) for $z\simeq6$ galaxies. 
Our further analyses were then based on the stacked images. The $J$ and $H$ 
bands do not cover \lya\ emission for our galaxies, so the morphology in the 
stacked images is purely from their UV continuum emission (other nebular lines 
can be safely ignored in general; see also Cai et al. 2011 and Kashikawa et 
al. 2012). In Figure 1, we show the thumbnail images of 44 (out of 67) 
galaxies that have more than $10\sigma$ detections of their total fluxes in 
the stacked images. We will focus on these 44 galaxies in this section.
Note that we excluded object no. 12, since it overlaps with a bright
foreground star, as explained in Paper I.

We do not show in Figure 1 the galaxies with $<10\sigma$ detections. Most of 
them are very faint. The others have shallower images (e.g., 1-orbit depth). 
We would not obtain reliable morphological information for these individual
faint galaxies, so we combined their images and made a single stacked image. 
The individual images were scaled before stacking, so that the galaxies all 
have the same magnitude. The final stacked image has a much higher S/N ratio.

\subsection{Size}

We use half-light radius $r_{\rm hl}$ to describe the size of a galaxy. 
The half-light radius $r_{\rm hl}$ is a radius enclosing a half of the total 
light. In Figure 2, the upper panel shows the measured $r_{\rm hl}$ as the 
function of $M_{1500}$, the absolute AB magnitude of the continuum at 
rest-frame 1500 \AA. The measured $r_{\rm hl}$ were calculated from
elliptical apertures using {\tt SExtractor} \citep{ber96}. 
Physical quantities such as $M_{1500}$ were 
derived in Paper I. The blue and red circles in Figure 2 represent the LAEs at 
$z\simeq5.7$ and 6.5 (including $z\simeq7$), respectively, and the green 
circles represent the LBGs at $z\simeq6$. We do not include object no. 67, 
since it has three well separated components and its $r_{\rm hl}$ is likely 
meaningless. The black star indicates the stacked image of the faint galaxies
described above. The measurement uncertainties in the upper panel were derived 
from the simulations described below. These galaxies roughly span a luminosity 
range of $-22 \le M_{1500} \le -19.5$, and a radius range of 
$0\farcs1 \le r_{\rm hl} \le 0\farcs3$ (or 0.6--1.7 kpc) without correction 
for PSF broadening. The median $r_{\rm hl}$ value is $0\farcs19$ ($\sim1.1$ 
kpc). Note that the sample in Figure 2 is dominated by the galaxies at
$M_{1500} \le -20$ mag. The stacked image of the faint galaxies 
has a relatively smaller radius 
$r_{\rm hl}\simeq0\farcs14$ ($\sim0.8$ kpc). The median $r_{\rm hl}$ value for 
the whole sample of 67 galaxies is $0\farcs16$ ($\sim0.9$ kpc), if we assume 
that the faint galaxies with $M_{1500} \ge -19.5$ mag in this sample have
$r_{\rm hl}$ smaller than this median value.

We estimated measurement uncertainties for $r_{\rm hl}$ from simulations. For 
each galaxy, we made a model galaxy using the two-dimensional fitting 
algorithm {\tt GALFIT} \citep{pen02}. A single component with a S\'{e}rsic 
function was fitted to the galaxy. This noiseless mock galaxy was put back at 
$300\sim400$ random positions (one position at a time) in the blank regions of 
the same science image. We then measured $r_{\rm hl}$ using the same way as we 
did for our real $z\ge6$ galaxies. We denoted the standard deviation of the 
measured $r_{\rm hl}$ as the measurement uncertainty of $r_{\rm hl}$ for this 
galaxy. The measurement uncertainties of $r_{\rm hl}$ in our sample (shown as 
the error bars in the upper panel of Figure 2) have a median value of 10\%.

We estimated intrinsic half-light radius $r_{\rm hl,in}$ and systematic
uncertainties from simulations as well. Because of the PSF broadening and 
lower surface brightness at larger sizes, the observed (or measured) 
$r_{\rm hl}$ is a complex function of galaxy size, brightness, and intrinsic 
profile. Simulations have been widely used to investigate these systematic 
effects \citep[e.g.][]{driver05,hau07,cib12,gra12,van12}.
We started with the {\tt GALFIT} parameters obtained above, and considered 
four parameters here, including the S\'{e}rsic index $n$, the axis ratio 
$b/a$, brightness, and $r_{\rm hl,in}$. Based on these parameters, we produced 
a large set of mock galaxies in a grid of $n$, $b/a$, $M_{1500}$, and 
$r_{\rm hl,in}$. The values of $n$ were chosen to be 1, 1.5, and 2 times the 
measured $n$ from our galaxies, and the values of $b/a$ were chosen to be 1, 
2/3, 1/3 times the measured $b/a$ from our galaxies. This is because the PSF 
broadening could largely decrease $n$ and increase $b/a$ in low-resolution 
images. The luminosity coverage ($-22.4\le M_{1500} \le -19.4$) and 
$r_{\rm hl,in}$ coverage ($0\farcs03 \le r_{\rm hl,in} \le 0\farcs4$) that we 
chose are roughly consistent with the actual coverage of our galaxy sample.
The mock galaxies were oversampled so that their $r_{\rm hl,in}$ sizes were 
at least 20 pixels. Then they were convolved with PSF images and rebinned
to match the pixel scales of our $HST$ images. Finally, each of the rebinned
mock galaxies was placed at many ($>300$) random positions in the blank 
regions of our science images. The $r_{\rm hl}$ of this mock galaxy 
is the median $r_{\rm hl}$ value measured at these random positions.

Figure 3 shows part of our simulation results. It illustrates the measured
$r_{\rm hl}$ as a function of $r_{\rm hl,in}$ at four different magnitudes.
It clearly shows that at small sizes, $r_{\rm hl}$ is significantly larger
than $r_{\rm hl,in}$ due to the PSF broadening. At large sizes, 
however, $r_{\rm hl}$ starts to fall short of
$r_{\rm hl,in}$, because we start to loose low surface brightness pixels at 
large sizes by detection. This happens at smaller sizes for fainter galaxies. 
At the faintest magnitude $M_{1500}=-19.5$ mag, $r_{\rm hl}$ will never 
exceed $0\farcs3$ in our images, regardless of $r_{\rm hl,in}$. 
On the other hand, if faint galaxies are always small, as seen in deeper $HST$ 
images \citep[e.g.,][]{win08,oes10,gra12}, their $r_{\rm hl}$ should not be
significantly smaller than $r_{\rm hl,in}$ (e.g. $r \le 0\farcs2$ at 
$M_{1500}=-19.5$ mag in Figure 3). Note that Figure 3 is similar to Figure 4 
in \citet{gra12}. Another simple way to correct for PSF broadening is to 
estimate $r_{\rm hl,in}$ in quadrature, i.e.,
$r^2_{\rm hl,in} = r^2_{\rm hl} - r^2_{\rm PSF}$, where $r_{\rm PSF}$ is the
PSF radius. Figure 3 shows that this equation is a good approximation at small 
sizes and/or high luminosities (green dashed lines), but it underestimates
$r_{\rm hl,in}$ elsewhere.

We used the relations in Figure 3 to estimate intrinsic sizes and associated
systematic uncertainties for our galaxies. The results are shown in the lower
panel of Figure 2. The error bars include both measurement and systematic
uncertainties, with systematic uncertainties being the dominant factors for 
most galaxies. With this correction, the values of $r_{\rm hl,in}$ for our 
galaxies at $M_{1500} \le -19.5$ mag range from $\le 0\farcs05$ ($<0.3$ kpc) 
to $\sim0\farcs3$ ($\sim1.7$ kpc), with a median value of $0\farcs16$ 
($\sim0.9$ kpc). The $r_{\rm hl,in}$ for the stacked object is about 
$0\farcs09$. The median $r_{\rm hl,in}$ value for the whole sample of 67 
galaxies is $0\farcs13$ ($\sim0.7$ kpc), if we assume that the faint galaxies 
at $M_{1500} \ge -19.5$ mag in this sample have $r_{\rm hl,in}$ smaller than 
this median value.

The galaxy sizes in our sample roughly agree with those of high-redshift
LAEs and LBGs with similar luminosities in the previous literature. 
For example, \citet{pir07} found that the average $r_{\rm hl}$ for a sample of
luminous LAEs at $z\sim5$ is $0\farcs17$. \citet{tan09} found a median
$r_{\rm hl}$ of $0\farcs15$ for LAEs at $z\sim5.7$. In the \citet{hathi08} 
and \citet{con09} LBG samples of $z=4\sim6$ galaxies, the $r_{\rm hl}$ ranges 
are $0\farcs1\sim0\farcs3$, similar to the $r_{\rm hl}$ range in our 
sample. Previous studies have shown that the galaxy size roughly scales with 
redshift as $(1+z)^{-m}$, with $m$ close to 1.1--1.2 
\citep[e.g.][]{fer04,bou06,oes10,mos12}, 
so the size of galaxies evolves slowly at 
high redshift. This is the reason that high-redshift galaxies have a similar 
size range. \citet{mal12} found, however, that LAEs have a roughly
constant size in the redshift range of $2.25<z<6$, and do not show a
size-redshift relation. While our sample does not have a large redshift
coverage, our galaxy sizes are well consistent with those in their sample.

\subsubsection{Size-luminosity relation}

Figure 2 shows that brighter objects appear to have larger $r_{\rm hl}$, 
meaning that more luminous galaxies tend to have larger physical sizes. This 
size-luminosity relation has been reported for both low- and high-redshift
star-forming galaxies \citep[e.g.,][]{tan09,oes10,gra12,ono12}. For example,
with a large sample photometrically-selected LBGs at $z\sim7$ in the CANDELS 
fields, \citet{gra12} found a strong relation $r_{\rm hl} \propto L^{\alpha}$, 
with slope $\alpha\simeq1/2$. In Figure 2, we illustrate the size-luminosity
relation by displaying the best log-linear fits (dashed lines). The best 
fitting results in the two panels are $r_{\rm hl} \propto L^{0.11 \pm 0.02}$ 
and $r_{\rm hl,in} \propto L^{0.14 \pm 0.03}$, respectively. 
Our slopes are much flatter than that in \citet{gra12} and those in other
fainter LBG samples. The reason is that our 
galaxy sample is much brighter. Our relation is derived from galaxies in the 
luminosity range of $M_{1500}\le -19.5$ mag, while the \citet{gra12} sample 
covers a range of $M_{\rm UV}\le -18$ mag. Their relation largely depends on 
the galaxies fainter than --19.5 mag, as seen in Figure 9 of their paper. In 
the brighter galaxies, their $r_{\rm hl}$ (or $r_{\rm hl,in}$) shows less of a 
trend with luminosity, as also pointed out by \citet{gra12}. In fact, our best
fit to the galaxies with $M_{1500}\le -20$ mag (dash-dotted line in Figure 2)
gives a nearly flat slope of $\alpha=-0.06 \pm 0.03$, suggesting little 
correlation between size and luminosity in the most luminous galaxies. 

It should be pointed out that the relation between the measured size and
luminosity could be affected by systematic effects shown in Figure 3. Figure 4 
illustrates how such effects shape the $r_{\rm hl}$-$M_{1500}$ relation for 
mock galaxies from the simulations above \citep[see also e.g.,][]{cib12}. 
The open circles indicate the intrinsic sizes $r_{\rm hl,in}$, and the filled 
circles are the measured sizes $r_{\rm hl}$ at different luminosities. The two 
colors red and blue indicate two different $HST$ images that the mock galaxies 
are placed in. The two panels are for two sets of S\'{e}rsic index $n$: 1 and 
2 times the measured $n$ from our $z\ge6$ galaxies. Figure 4 shows that 
fainter galaxies (with the same intrinsic size)
appear to be smaller, as already shown in Figure 3. For 
the same reason, larger galaxies (with the same intrinsic luminosity) appear 
to be slightly fainter. We have taken into account these effects in Figure 2.
These effects do not have significant impact on the size-luminosity relation 
in Figure 2: the corrected relation does not become flatter, because this
relation also depends on other factors such as the source distribution.

Finally, we point out that the size-luminosity relation in our sample is not 
affected by a possible selection effect, i.e., that we may have missed some 
faint galaxies with large sizes during galaxy candidate selection. The reason 
is that these galaxies were selected in optical images. They are bright and 
point-like sources in the optical broad bands (for LBGs) or narrow bands (for 
LAEs). Also note that the exclusion of faint galaxies in our sample does not 
introduce bias to our results. Our $HST$ near-IR data have relatively uniform 
depth (two orbits per band per pointing), so the $10\sigma$ cut indeed puts a 
flux limit on $M_{1500}$ in Figure 2, which does not affect our results in the 
bright region of $M_{1500}<-19.5$ mag.

\subsection{Nonparametric Measurements of Morphology}

In this subsection, we will characterize galaxy structure and morphology using 
nonparametric methods, including the $CAS$ (Concentration, Asymmetry, and 
Smoothness) system \citep{con03}, the Gini coefficient $G$, and the $M_{20}$ 
parameters \citep{lot04}. 
These methods have been widely used for lower-redshift galaxies. 
They usually provide reliable description of galaxy structure, and are able to 
distinguish different types of galaxies. To obtain accurate measurements of 
these quantities, two criteria are often required: high S/N ratios and large 
object sizes compared to the PSF size. For low-redshift galaxies, especially 
those in $HST$ images, the sizes of galaxies are many times larger than PSF, 
so the two criteria are naturally met in deep $HST$ images. At higher 
redshift, galaxies are fainter and smaller, so it is difficult to meet the two 
criteria, and these parameters become less reliable. We investigate the 
reliability of these measurements for our sample in detail below.

\subsubsection{$CAS$, Gini, and $M_{20}$ parameters}

Concentration ($C$) measures how compact the galaxy light profile is.
We adopted the commonly used definition $C=5\,{\rm log}\,(r_{80}/r_{20})$ 
\citep[e.g.,][]{ber00,con03}, where $r_{80}$ and $r_{20}$ are the radii that 
contain 80\% and 20\% of the total galaxy flux, respectively.
Asymmetry ($A$) measures how rotationally symmetric a galaxy is
\citep{abr96,con03}. It is calculated by subtracting the image rotated by 
180$\degr$ from the original galaxy image. 
Smoothness ($S$) or clumpiness measures how clumpy a galaxy is \citep{con03}.
\citet{con09} found that $S$ fails to well describe clumpiness for $z=4\sim6$
galaxies, so we did not calculate $S$ in this paper.
The $M_{20}$ parameter, or the second-order moment of the brightest 20\% of 
the galaxy, is similar to the concentration $C$, and measures how the galaxy 
light is concentrated \citep{lot04}. The Gini coefficient ($G$) describes how 
even the galaxy light distribution is \citep{abr03,lot04}. We computed $G$ and 
$M_{20}$ as described by \citet{lot04}.

The measurements of these parameters are shown in Figure 5. The measurement 
uncertainties were estimated from simulations using the same method as we did 
for the uncertainties of $r_{\rm hl}$. We took the model galaxies obtained in
Section 3.1, and put them at many random positions in the blank regions of
our $HST$ images. We then measured $CAGM_{20}$ at each position, and calculated
the standard deviations of these parameters. The standard deviations, or
measurement uncertainties, are shown as the error bars in Figure 5.
These uncertainties include the effects of S/N in the images, but do not 
account for systematic uncertainties associated with sparse spatial sampling
of distant small and faint sources which we investigate in Section 3.2.2.

Compared to low-redshift galaxies, our galaxies are located in a much narrower 
range in the parameter space (Figure 5; see also Figures 6 and 7). They appear
to be less concentrated and more asymmetric, and their light distribution is 
more even. This is likely because the 
measured quantities have been substantially affected by the low resolution of 
our images. For example, $C$ has a narrow range between $\sim$2 and $\sim$3, 
and there is a lack of highly-concentrated values of $C$. Due to the low 
spatial resolution, the measurements of $r_{80}$ and $r_{20}$ are not robust. 
In particular, for highly-concentrated galaxies, the inner radius $r_{20}$ is 
smaller than one pixel and is likely significantly overestimated, so $C$ is 
underestimated. We will discuss these systematic effects using simulations 
below.

\subsubsection{Systematic effects from simulations}

In order to address how the low spatial resolution of the images affect the 
measurements of the parameters $CAGM_{20}$, or how reliable these parameters 
are (for $z\ge6$ galaxies), we ran a series of slightly different simulations 
than we did in Section 3.1. Here we started with low-redshift real galaxies 
instead of model galaxies, because model galaxies are smooth, and do not cover 
a large range of the parameter space. For example, $A$ is zero for a noiseless
single-component model galaxy. We chose to use the galaxy images in the 
library of galSVM \citep{hue08,hue11}. These low-redshift galaxies are large 
and bright, and their images have high S/N ratios. We randomly selected 1/10 
of the galaxies from this library. We then visually inspected these galaxies 
and removed those with possible foreground stars. Our final sample consists of 
740 galaxies. Their morphological parameters are shown in Figure 6
(the scale in this figure is very different from that in Figure 5).
These galaxies cover a large range of the parameter space of $CAGM_{20}$.

By rescaling these real galaxies in size and flux, 
we produced a large set of mock galaxies with
high spatial resolution at $z\ge6$, in the grid of magnitude 
($-22<M_{1500}<-19$) and size ($0\farcs05 \le r_{\rm hl,in} \le 0\farcs4$). 
The mock galaxies were then convolved with the PSF images and rebinned to 
match the pixel scales of our $HST$ images. Finally, the rebinned mock 
galaxies were placed at many random positions in the blank regions of the 
$HST$ images, and their morphological parameters were measured. 
Figure 7 shows an example of our simulation results (black dots). In this 
example, we preserved the relative magnitudes and sizes of the galaxies, and 
scaled the sample as a whole so that the scaled sample covers a similar range 
of magnitude and size as our $z\ge6$ galaxies do. This is the best way to 
preserve the scatter of source distributions in the parameter space. 

The comparison between Figure 6 and Figure 7 shows that the parameters of 
these mock galaxies measured in our low-resolution images are quite different 
from the intrinsic values: they occupy a smaller range of the parameter space. 
For example, the mock galaxies are significantly less concentrated in our 
images ($2<C<3$ in Figure 7 $vs.$ $2<C<5$ in Figure 6). As we already 
mentioned, $C$ is significantly underestimated for highly-concentrated 
galaxies because $r_{20}$ is much smaller than one pixel in our $HST$ images.
Since $M_{20}$ also describes galaxy concentration, it is underestimated as
well (less negative here). This was also noticed by \citet{lot06}.
Due to the low spatial resolution, our images cannot resolve subtle 
structures, so the mock galaxies show lower $G$ coefficients, or more evenly 
distributed light. The limitation of $G$ has been reported
\citep[e.g.][]{lisker08}. Also because of the low resolution, the $A$
values of the mock galaxies are much larger.

In Figure 7 we also plot our $z\ge6$ galaxies (red circles). Their 
morphological measurements are directly taken from Figure 5. Their positions 
in the parameter space are quite consistent with those of the mock galaxies.
This suggests that the $z\ge6$ galaxies are possibly not intrinsically less 
concentrated, more asymmetric, or less even in light distribution. It is 
simply because our images do not meet one critical requirement to use these
methods, i.e., large galaxy sizes compared to the PSF size, so our 
measurements have been systematically biased by the low-resolution images.
On the other hand, these parameters are probably still meaningful for galaxies 
at similar redshifts, if they are measured in the $HST$ images of the same 
depth and pixel size, i.e., for example, intrinsically more concentrated 
galaxies are still more concentrated in our $HST$ images as measured by $C$. 
We demonstrate this using the simulations of the 740 mock galaxies shown in 
Figure 7. We first choose the galaxy pairs whose difference of the 
{\it measured} $C$ (or any of $CASM_{20}$) are larger than $2\sigma$ ($\sigma$ 
is the measurement uncertainty). Among these galaxy pairs, we further select 
the pairs in which the galaxy has larger {\it intrinsic} $C$ than the other 
one still has larger {\it measured} $C$ in the low-resolution images. The 
fraction of such pairs for $C$ is 93\%. The fractions for $ASM_{20}$ are 70\%, 
98\%, and 92\%, respectively. If we increase $2\sigma$ to $3\sigma$ above, the 
fractions for $CASM_{20}$ increase to 97\%, 78\%, 100\%, and 95\%, 
respectively. This suggests that for the vast majority of galaxies, the low 
resolution of our $HST$ images does not change their relative values of these 
morphological parameters. We emphasize that the {\it absolute} measured
values of these parameters for $z\ge6$ galaxies in $HST$ images are not to be 
compared to the {\it absolute} measured parameters of lower-redshift galaxies.

\subsubsection{Relations among the morphological parameters}

The morphological parameters correlate with each other. We have seen such 
correlations in low-redshift galaxies \citep[e.g.][]{lot04,lot06,con09}. 
In Figure 6, the dashed lines are the best linear fits to the 740 low-redshift 
galaxies, and show the relations among $CAGM_{20}$. As expected, these 
parameters are correlated with each other. The moment $M_{20}$ is correlated 
well with $C$, $A$, and $G$. Both $M_{20}$ and $C$ describe how the galaxy 
light is concentrated, and they are strongly correlated by definition. 
The relations between $M_{20}$ and $A$ \& $G$ reflect that more concentrated 
galaxies have more rotationally symmetric profiles and more unevenly 
distributed light. The Gini coefficient $G$ is correlated with $C$ and $A$, 
in addition to $M_{20}$. Its relations with $C$ and $A$ indicate that galaxies 
with more concentrated or more rotationally symmetric profiles tend to have 
more unevenly distributed light. These relations still exist among the 740 
mock galaxies in Figure 7, since the relative values of the parameters are 
preserved as we discussed above. On the other hand, some relations become 
weaker with larger scatter. For example, the relations between $C$ and $AG$ 
are much flatter, mainly because the coverage of $C$ has shifted from 
$2 \le C\le 5$ to $2 \le C \le 3$. 

In Figure 5, the red dashed lines show the correlations of $CAGM_{20}$ for
our $z\ge6$ galaxies. These correlations are consistent with those for
the mock galaxies in Figure 7. In particular, the relations between $M_{20}$
and $CAG$ are still fairly well preserved, and the relations between $C$ and 
$AG$ are as weak as shown in Figure 7.
We will see in the next subsection that these corrections for relatively 
bright galaxies in our sample are better with smaller scatter (Figure 10).
Therefore, although our measured $CAGM_{20}$ of the $z\ge6$ galaxies are
systematically biased by the low-resolution images, 
these parameters, when interpreted carefully, are still somewhat meaningful 
for galaxies measured in the $HST$ images of the same depth and pixel size. 

\subsection{Interacting Systems}

One of the interesting morphological topics is to study interacting/merging
systems, which traces hierarchical mass assembly in the CDM scenario.
A close visual inspection of Figure 1 shows that some galaxies are clearly 
extended with interacting or multi-component features. At lower redshift, 
these systems can be identified by pair counts \citep[e.g.,][]{lef00}, the 
$CAS$ system \citep[e.g.,][]{con03}, or the $G$ and $M_{20}$ parameters 
\citep[e.g.,][]{lot08}. Galaxies at $z\ge6$ are faint and small, so it is 
difficult to properly distinguish regular and interacting/merging systems. 
Although we have derived morphological parameters $CAGM_{20}$, they were 
biased, and their ability to identify merging systems at $z\ge6$ has never 
been examined. Therefore, we identify interacting/merging systems by visual 
inspection. Visual classification was the earliest way for galaxy 
classification, and in many cases is still the best way to identify merging 
systems at high redshift.

We considered the following two types of galaxies as candidate 
interacting/merging systems: 1) galaxies with two or more distinct cores; and 
2) galaxies with extended/elongated features and/or long tails. We identified
these systems in 24 relatively bright ($M_{1500}\le-20.5$ mag) galaxies. 
For fainter galaxies, our images are not deep enough to properly identify all 
faint components or extended features. Figures 8 and 9 show 12 galaxies that 
were identified as interacting systems with $M_{1500}\le-20.5$ mag. Figure 8 
shows 11 galaxies in the stacked ($J+H$) images, and Figure 9 shows the 
$z=6.96$ LAE (no. 62) in the two $J$ bands. The red profiles are the surface 
brightness (SB) contours of the 
rest-frame UV emission. Six of them clear show double or multiple clumps, 
including no. 4, 24, 34, 49, 62, and 67. They usually have one bright core and 
one or more fainter clumps. No. 62 and 67 are particularly interesting. 
No. 62 in Figure 9 has almost two identical components, and no. 67 has three 
widely-separated cores. More discussion is given in the next subsection. 
Another 3 galaxies, including no. 36, 58, and 61, do not clearly show multiple 
clumps, but have long tails like tidal tails seen in low-redshift merging
galaxies. The rest of the 12 systems (no. 15, 44, and 47) do not show multiple
components or tails, but they are rather extended and elongated.
They could be in the end of the merging process.

We estimate the fraction of mergers among galaxies with $M_{1500}\le -20.5$ 
mag. The fraction is 50\%, or 38\% if we exclude the three galaxies that do 
not show multiple components or tails. The fraction is even higher in the 
galaxies with $M_{1500}\le -21$ mag. We have 18 galaxies in this magnitude 
range, and 10 of them are mergers. The fraction of mergers is 56\%, or 39\% if 
we exclude the three galaxies mentioned above. This is consistent with the 
fractions in the brightest galaxies at low redshift of $z\simeq2-3$. 
For example, the fraction of mergers in the $M_B<-21$ mag galaxy sample of 
\citet{con03} is 40\%--50\%. The typical fraction in the brightest galaxies 
in the sample of \citet{law12} is also $\sim40$\%. The merger fraction in 
fainter galaxies is smaller, because merger systems consist of multiple 
components, and usually have stronger SFRs and UV emission. 
For the galaxies fainter than $M_{1500}\simeq-20.5$ mag in our sample, 
our images are deep enough to identify double-core systems with comparable 
emission. We find that these systems are very rare at $M_{1500} \ge -20.5$ 
mag. Studies in deeper fields have indicated a low merger fraction in 
low-luminosity galaxies. For example, \citet{oes10} presented the morphology 
of 16 $z\ge7$ LBGs in the HUDF, and only found two galaxies with extended 
features. 

\subsubsection{Notes on individual objects}

{\it Galaxy no. 62.}
No. 62 is a $z=6.964$ LAE. It is the first spectroscopically-confirmed LAE 
at $z\sim7$ \citep{iye06}. Figure 9 shows that it has two similar components 
in the both $J$ bands. We used {\tt GALFIT} to model the two components.
Two S\'{e}rsic functions were fitted to the two components (two left-hand
images in Figure 9) simultaneously. The middle images in Figure 9 show the best
model fits, and the residuals are on the right-hand side. The two components
in the both bands can be well described by the S\'{e}rsic function.
The separation between the two cores is about $0\farcs2$ ($\sim1$ kpc)
\citep[see also][]{cai11}.

{\it Galaxy no. 67.}
No. 67 is a LAE at $z=6.595$. It was discovered as a giant LAE by
\citet{ouc09}. It is one of the brightest galaxies in terms of both \lya\
luminosity and UV continuum luminosity. The most striking feature is the
three well separated cores lined up. The central core is relatively weak.
This galaxy is clearly resolved in our ground-based $z-$ and $y$-band images.
The separate between the two side cores reaches $\simeq1\farcs2$, or 7 kpc.
This is the largest separation we have seen at $z\ge6$. This object also
has strong emission in the IRAC bands, suggesting a large stellar mass of
$\ge10^{10}M_{\sun}$ \citep{ouc09}.

\subsubsection{Morphology of interacting systems}

As we mentioned, the above interacting galaxies were identified with visual
inspection. Morphological parameters were not used for selection
because the criteria of
interacting systems at $z\ge6$ are unclear. We check our interacting systems 
in the parameter space in Figure 10. Figure 10 is similar to Figure 5, but only
plots the galaxies with $M_{1500}\le -20.5$ mag in our sample. The red squares 
represent the interacting galaxies. Figure 10 shows that the interacting 
galaxies are almost indistinguishable from the rest in the parameter space.
Asymmetry $A$ is widely accepted as an efficient parameter to identify mergers 
at low redshift \citep[e.g.][]{con03,con09,lot06,lot08,law12}.
In Figure 10, however, $A$ is not a good indicator of mergers any more,
although the galaxies with the largest $A$ in our sample are mostly interacting 
systems.

An interesting feature in Figure 10 is the better correlations among the 
morphological parameters compared to Figure 5. The scatter in the relations 
is smaller for these bright galaxies with $M_{1500}\le -20.5$ mag. 
There are two explanations. One is that bright galaxies have higher
S/N ratios and cover more pixels in the $HST$ images, so their morphological
measurements are more robust. The other one is that galaxies with different
luminosities occupy slightly different parameter space, so a sample covering
a smaller luminosity range shows a smaller scatter in Figure 10. 
The real reason is very likely the combination of the two. In any case, 
Figure 10 strengthens our earlier conclusion on the existence of strong
correlations among $CAGM_{20}$ for our $z\ge6$ galaxies.

\section{\lya\ MORPHOLOGY}

In this section we will study the \lya\ morphology of LAEs using our 
ground-based narrow-band images. We will not measure structure and morphology 
for individual galaxies, since they are mostly point-like objects in the 
ground-based images. Although these images have excellent PSF sizes of 
$\sim0\farcs5-0\farcs7$, the PSF sizes are still much larger those of the 
$HST$ images. Therefore, we will focus on \lya\ halos around LAEs, which could 
extend many arcsec from the objects. We will also compare the positions of 
\lya\ emission with those of UV continuum emission, and find any possible 
positional difference between the two.

\subsection{\lya\ Halos}

Because of the resonant scattering of \lya\ photons by neutral hydrogen, \lya\ 
emission could form large diffuse \lya\ halos around high-redshift galaxies. 
\citet{ste11} first found very extended \lya\ halos in a sample of luminous 
galaxies at $2<z<3$. The galaxies were UV continuum-selected, but more than a 
half of them show net \lya\ emission and $\sim20$\% have \lya\ EW greater than 
20 \AA. They were able to find large \lya\ emission halos ($\ge80$ kpc) in the
stacked images of all sub-samples of their galaxies. They further claimed that 
all LBGs would be classified as LAEs or \lya\ blobs, if imaging data are deep
enough to detect \lya\ halos. \citet{mat12} confirmed the existence of 
extended \lya\ halos around $z\simeq3$ galaxies. They used more than 2000 LAEs 
at $z\simeq3.1$, and grouped them into sub-samples based on luminosity and 
surface overdensity. They stacked narrow-band (\lya) images for each 
sub-sample, and found that all stacked images show extended ($>60$ kpc) \lya\ 
emission halos. 
Recently \citet{fel13} found that the existence of \lya\ halos around LAEs is
not convincing. They also used a large sample of a few hundred LAEs at 
$z\simeq2.1$ and 3.1. They paid particular attention to systematic effects 
from large-radius PSF and large-scale flat fielding, etc. When these effects
were taken into account, they did not find strong evidence of extended \lya\ 
halos in the stacked narrow-band images at either redshift. They tried a few 
ways to reconcile the discrepancy between their results and the previous 
results, yet the reason of the discrepancy is still not clear.

Stacking of narrow-band images has not been done for $z\ge6$ galaxies. On the 
other hand, cosmological simulations have predicted the existence of extended 
\lya\ emission around $z\ge6$ galaxies \citep[e.g.,][]{zheng11,dij12,jee12}. 
For example, by including the resonant scattering of \lya\ photons in both 
circumgalactic media and intergalactic medium (IGM), \citet{zheng11} showed 
that the \lya\ emitting halo in a high-redshift galaxy can extend up to 1 Mpc. 
They further pointed out that such halos could be detected by stacking 100 
$z\simeq5.7$ LAEs in 4-hr exposure narrow-band images in the SXDS.
Here we combine the narrow-band (\lya) images of LAEs at $z\simeq5.7$ and 6.5.

In order to detect diffuse \lya\ halos around LAEs, we made use of all the 
known LAEs at $z\simeq5.7$ and 6.5 in the SDF and SXDS 
fields from \citet{kas11} and \citet{ouc08,ouc10}.
The narrow-band images were taken with the Subaru telescope. 
The total integration time in the NB816 and NB921 bands are
10 and 15 hr for the SDF \citep{kas04}, and $\sim$4 hr and $\sim$10 hr (depth
slightly varies among five pointings) for the SXDS \citep{ouc08,ouc10}.
The data reduction of the SDF images were presented in Section 2 of Paper I.
The reduction of the SXDS images were done in the same way. These narrow-band
images have great depth with excellent PSF sizes of $0\farcs5\sim0\farcs7$.
We rejected a small number of galaxies that are either very faint or blended
with nearby bright objects. The final sample contains 43 LAEs at $z\simeq5.7$
and 40 LAEs at $z\simeq6.5$.

To stack these images, we first cut image stamps for all individual objects. 
We then re-sampled image stamps, so that the objects are in the centers of 
the images. After all other objects in the images were masked out, we co-added 
(average) the images with sigma-clipping ($5\sigma$ rejection). Figure 11 
shows the stacked narrow-band \lya\ images at the two redshifts. We did not 
subtract away continuum images. As we showed in Paper I, our LAEs have \lya\ 
EWs greater than 20 \AA. The median EW value is 80 \AA, so their narrow-band 
photometry is completely dominated by the \lya\ emission. Subtracting 
continuum images would significantly increase the background noise. 
In the above procedure we did not scale the objects to the same magnitude 
either, since we do not know how the SB of the \lya\ halos scales with the 
total \lya\ flux in a LAE. We performed a series of tests by stacking the 
images in different ways, including image stacking with median, with objects 
scaled to the same magnitude, and with bright objects only. The results are 
all very similar to the images shown in Figure 11. Our stacked images of LAEs 
reach a depth of roughly
$\rm 1.2\times 10^{-19}\ erg\,s^{-1}\,cm^{-2}\,arcsec^{-2}$ ($1\sigma$) 
in both bands (see also Figure 12). We should point out that this depth is 
shallower than what we predicted assuming that the noise goes down with the 
square root of the image number. The reason is that the input images are not 
purely blank images, whose noise is dominated by background Poisson noise. 
These images are extremely deep and crowded, so LAEs have many nearby objects. 
Although we were able to mask out the bright pixels of the nearby objects, 
we could not remove the light outside these pixels that may extend many pixels 
away. When we combine many images, these nearby objects contribute significant 
unavoidable noise to the co-added images.

Figure 11 clearly shows that the stacked \lya\ emission is compact at the 
both redshifts, and do not show extended diffuse halos. As a reference,
we also show in Figure 11 the stacked images of stars (point sources). 
The stars are chosen to be separated, bright, but not saturated, in the same 
narrow-band images. They are combined in the same way as we did for LAEs. 
Figure 12 shows the radial profiles of the stacked LAEs (solid lines with
error bars) as well as the profiles of the stars (dotted lines). The \lya\ 
profiles roughly follow the PSF profiles, and their SB reach zero at 
$r\ge2\arcsec$.
In the upper panel, the profile FWHM of the stacked $z\simeq5.7$ LAE and star 
are $0\farcs49$ and $0\farcs62$, respectively. In the lower panel the two FWHM 
values are $0\farcs61$ for the $z\simeq6.5$ LAE and $0\farcs77$ for the star. 
The LAE profiles are broader than the PSF sizes by $\sim26$\%, and exhibit 
slightly longer tails than the PSF profiles do. This is simply because 
galaxies are not point sources, and indicates that the \lya\ emission is
resolved but not very extended. In the previous sections we show that our 
LAEs have a range of sizes in their rest-frame UV emission. \citet{hathi08}
stacked broad-band HUDF images for $z\sim4-6$ LBGs and found that the galaxy
SB profiles are apparently broader than the PSF profiles. \citet{fin11} found
that two $z\simeq4.4$ LAEs have large sizes in \lya\ than in UV continuum.
So the slightly broader \lya\ SB profiles compared to the PSF are just the 
nature of galaxies, and are not likely caused by diffuse 
\lya\ halos predicted or found in previous studies. 
The $z=5.7$ profile seems to exhibit slightly more 
extended radius than its PSF profile. Given the $1\sigma$ limit of 
$\rm \sim1.2\times 10^{-19}\ erg\,s^{-1}\,cm^{-2}\,arcsec^{-2}$, its 
SB is still consistent with zero at $r\ge2\arcsec$. Therefore, Figure 12
does not show convincing evidence of extended \lya\ halos.

It is difficult to answer whether our stacked images are deep enough to detect
\lya\ halos if they do exist at the two redshifts. Our images are certainly
deep enough to detect the LBG halos at $z\simeq3$ reported by \citet{ste11} 
and the LAE halos at $z\simeq3.1$ in \citet{mat12}. 
The depth to detect $z\ge6$ LAE halos is 
observationally unknown. From cosmological simulations, \citet{zheng11} 
predicted \lya\ halo sizes in $z=5.7$ LAEs. They found two characteristic 
scales for the halos. The inner steeper one extends to $3\arcsec-4\arcsec$, 
and the outer flatter one extends to a few tens of arcsec. While our images 
are not deep enough to detect the outer halos, they are almost deep enough 
to detect the inner scale halos as seen in Figure 3 of \citet{zheng11}, where 
the \lya\ radial profiles are shown for LAEs in dark matter halos of 
$\sim10^{11}M_{\sun}$. As pointed out by \citet{zheng11}, the size of diffuse 
\lya\ emission also depends on the mass of dark matter halo. If the average 
mass of the dark matter halos in our LAEs is smaller than $10^{11}M_{\sun}$, 
our current data may not be able to detect the diffuse \lya\ emission. 

It is also likely that the \lya\ halos (if they exist) have been diluted to a 
much lower level during the construction of the stacked images. The stacked 
images can properly recover \lya\ halos only when halos are smoothly and 
symmetrically distributed around galaxies. If halos are highly asymmetric
and/or clumpy, the emission of halos will be significantly diluted in 
average stacked images, and could totally disappear in median stacked images.
From the observations of $z\sim6$ quasars or cosmological simulations,
we know that the distribution of IGM at $z\sim6$ is inhomogeneous 
\citep[e.g.][]{fan06,mes10}. If the distribution of IGM affects the 
shape of \lya\ halos (via resonant scattering), the distribution of \lya\ 
halos is also likely asymmetric and clumpy.

Finally, it is possible that these LAEs do not have extended \lya\ halos,
or that their halo emission is not as strong as predicted by \citet{zheng11}, 
especially when dust is taken into account \citep[e.g.][]{fin11}. 
\citet{zheng11} did not consider dust in their simulations. We know that 
high-redshift LAEs are not free of dust. In particular, the brightest galaxies 
may exhibit significant dust extinction, as implied by their UV colors (see 
Paper I). When \lya\ photons are resonantly scattered by dusty neutral 
hydrogen, the \lya\ emission is substantially reduced \citep{yajima12}. 
The reduction is more severe at 
larger distance from the object, because photons at larger distance need to 
pass through more dust before they escape. This process would significantly 
reduce the visibility of possible diffuse \lya\ emission, and makes it much 
more difficult to detect it. A much larger sample of LAEs is needed to answer 
this question.

\subsection{\lya-Continuum Misalignment}

The comparison between the positions of UV continuum and \lya\ emission
provides useful information on how \lya\ photons escape from a galaxy.
\lya\ and UV continuum photons are usually come from the same star-forming
regions, although \lya\ photons are likely more sensitive to the regions with
more recent star-forming activity. As we mentioned earlier, \lya\
emission is complicated by resonant scattering and IGM absorption. So the 
observed position of \lya\ emission could be different from the position of
UV continuum emission. For example, a large positional difference has been
found in a $z=3.334$ galaxy \citep{rauch11}.
Due to the small sizes of high-redshift galaxies,
current ground-based observations are not able to detect these positional
differences. We rely on $HST$, which has observed a large number of 
high-redshift galaxies. $HST$ observations were mostly made for rest-frame UV 
continuum emission, and there is usually no suitable $HST$ narrow-band filters 
for \lya\ emission. One example for the $HST$ imaging of \lya\ emission is the 
work of \citet{fin11}, who observed the \lya\ emission of a small sample of 
$\simeq4.4$ LAEs with a narrow-band filter. They did not find strong evidence
of positional misalignment between UV and \lya\ emission.

We used our large sample of LAEs to search for possible positional offsets
between UV and \lya\ emission at $z\simeq5.7$ and 6.5. We used our $HST$ 
images as UV continuum images and Subaru narrow-band images as \lya\ images. 
As mentioned above, the narrow-band images have excellent PSF FWHM sizes 
around $0\farcs5\sim0\farcs7$. During the construction of the $HST$ images 
(Paper I), we have matched the coordinates of the $HST$ images to those of the 
optical images. To avoid any large-scale variation, we refine the coordinates 
of the $HST$ images. For each LAE, we found 10--20 nearby objects that are 
relatively bright and round. We then matched the positions of the nearby 
objects in the two sets of images. The typical refinement was smaller than the 
size of one pixel ($0\farcs06$). The uncertainty in the object positions, 
derived from the distribution of the nearby objects, is about the size of 1--2 
pixels. We plot the \lya\ positions on top of the UV continuum positions.
Figure 13 shows a few examples of bright galaxies. The red profiles are the 
contours of the UV emission SB, and the green crosses indicate 
the positions (and $1\sigma$ uncertainties) of the \lya\ emission.

For the majority of the LAEs in our sample, the positions of the UV continuum 
and \lya\ emission agree with each other. In particular, we do not find 
positional misalignment at a significance level of $>2\sigma$ among almost
all compact and round LAEs. Object no. 3 in Figure 13 is a typical example, in 
which the center of the \lya\ emission is close to the center of the continuum 
emission. For the merging/interacting systems, however, we see significant 
positional differences. Figure 13 shows the examples of these systems. They 
exhibit a variety of \lya\ positions relative to the peak positions\footnote{
For an interacting galaxy, the position of its peak emission (as seen from the 
SB contour) could be very different from the position of the overall galaxy
emission measured by, for example, fitting Gaussian to the marginal x,y 
distributions (used by IRAF DAOPHOT). Here our positions refer to the 
positions of peak emission.}
of the UV continuum emission:
1) \lya\ positions close to the positions of the brightest components in the 
UV images, including galaxies no. 15, 47, and 61; 2) \lya\ positions close to 
the positions of the fainter components or merger tails in the UV images, 
including no. 14, 44, and 58; 3) \lya\ positions are somewhere between the 
positions of the bright and faint components, but closer to the bright 
components, including no. 4 and 49. No. 62 and 67 are again interesting. 
The \lya\ position of no. 62 is roughly in the middle of the two similarly
bright components. No. 67 does not show three distinct \lya\ emission clumps,
as its UV emission does. Instead, it shows
a single bright \lya\ emission core with some extended features. 
The \lya\ center is not at any of the three UV clumps. It is between the 
left and central clumps, and slightly closer to the central one.

Our results suggest that in compact galaxies the observed location of \lya\ 
emission does not deviate from its original position, while in 
merging/interacting systems, the observed \lya\ location could be 
significantly different from its original position, without preferential
positions of offsets. If the final location of \lya\ emission is determined by 
the process of resonant scattering, our results can be explained by the 
interstellar medium (ISM) distribution \citep[e.g.][]{fin11}. In a 
non-disturbed galaxy, the ISM distribution is relatively symmetric around the 
object (not necessarily uniform, it could be clumpy). The random scattering 
of photons does not have preferred directions, so the observed location is 
still close to its original location. In an interacting system, the ISM is 
re-distributed by merging activity. The distribution of the disturbed ISM is 
therefore not symmetric any more. This results in an offset of the observed 
\lya\ position.

This result could have an impact on spectroscopy of bright LBGs at $z\ge6$
\citep[e.g.][]{stark11,cur12}. While most of the positional offsets in our
sample are 
smaller than $0\farcs2$, at least two are around $0\farcs3-0\farcs4$.
If one uses a $\le1\arcsec$ slit to identify their \lya\ emission lines, based 
on the positions of continuum emission, one could miss them due to the large
offsets. However, glaxies with such large offsets are very rare, so this 
result will not largely reduce the success rate of identifying \lya\ lines
in LBGs.

\section{DISCUSSION AND SUMMARY}

The comparison in Section 3.1 shows that the sizes of our galaxies are
consistent with those of bright $z\ge6$ LAEs and LBGs in previous studies.
This disagrees with the claim of \citet{dow07} that LAEs are more compact than
LBGs. This is likely because the \citet{dow07} sample is very small while LAEs
have a large range of sizes. It is indeed difficult to make proper comparisons
without a large sample, as the galaxy size depends on redshift and luminosity.
Most of our galaxies are LAEs. Another large LAE sample is the $z\simeq5.7$
LAE sample by \citet{tan09}, who observed a number of LAEs at $z\simeq5.7$
with the $HST$ ACS F814W filter. Although this filter includes \lya\ emission,
it is so wide that its emission is dominated by UV continuum. They found that
the average size of the sample is $0\farcs15$, similar to ours. Among 
photometrically-selected LBG samples at $z\ge6$, a recent large sample is the
sample of \citet{gra12}. This sample contains a number of bright (as well as
faint) LBGs at $z\sim6-7$ in the CANDELS field. 
The sizes of these galaxies are well consistent with those of our galaxies at 
the same luminosities. Therefore, we conclude that LAEs and LBGs with similar 
luminosities have similar physical sizes.

In this paper (and Paper I), galaxies found by the narrow-band technique 
are defined as LAEs and those found by the dropout technique are defined as
LBGs. As we discussed in Paper I, this classification only reflects the 
methodology that we apply to select galaxies. It does not mean that the two
types of galaxies are intrinsically different. In Section 5.3 of Paper I,
when we derived the UV continuum luminosity function of LAEs, we used 
another popular definition of LAEs based on the \lya\ EW, i.e., a galaxy is 
a LAE if its \lya\ EW is greater than 20 \AA. With this definition, almost all 
the galaxies in our sample are LAEs. This definition is physically more 
meaningful, but observationally difficult, because one can easily obtain a 
flux-limited sample, not a EW-limited sample. We have 16 LBGs (former 
definition) in our sample. They are not typical LBGs. They are 
spectroscopically confirmed, and thus only represent those with strong \lya\ 
emission. In Paper I we found that our LAEs and LBGs are indistinguishable in 
many aspects of the \lya\ and UV continuum properties.
In this paper, we further found that these LAEs and LBGs have similar physical 
sizes and morphological parameters. This confirms one of our conclusions in 
Paper I that LAEs are a subset of LBGs with strong \lya\ emission lines.

This paper is the second in a series presenting the physical properties of
a large sample of spectroscopically-confirmed galaxies at $z\ge6$. The sample 
consists of 51 LAEs and 16 LBGs, and represents the most luminous galaxies in
terms of \lya\ luminosity (for LAEs) or UV continuum luminosity (for LBGs)
in this redshift range. In Paper I we derived various properties of rest-frame
UV continuum and \lya\ emission. In this paper we have conducted a detailed 
structural and morphological study of the galaxies using deep $HST$ near-IR 
images and Subaru narrow-band images. In order to measure the morphology of 
rest-frame UV continuum emission, we constructed a stacked $HST$ image for 
each object by combining its $J$- and $H$-band images. UV morphology was then 
measured for those with $>10\sigma$ detections (roughly corresponding to
$M_{1500}\le-19.5$ mag) in the stacked $HST$ images.

We have used half-light radius to describe the sizes of galaxies. 
The intrinsic sizes $r_{\rm hl,in}$ (at $M_{1500}\le-19.5$ mag) in our sample, 
after correction for PSF 
broadening, are from $\le 0\farcs05$ ($\le0.3$ kpc) to $\sim0\farcs3$ 
($\sim1.7$ kpc), with a median value of $0\farcs16$ ($\sim$0.9 kpc). These
values are consistent with those of bright, photometrically-selected LBGs at 
similar redshifts. 
Additionally, more luminous galaxies tend to have larger sizes,
exhibiting a weak size-luminosity relation $r_{\rm hl,in} \propto L^{0.14}$. 
The slope 0.14 is significantly flatter than those in fainter LBG samples.
Our objects show a wide range of morphology in the $HST$ images, including
compact galaxies and double/multiple component systems. The brightest galaxies
in the sample have a large fraction of merging/interacting systems. The 
fraction of mergers reaches $40-50$\% at $M_{1500}\le-20.5$ mag.

We have tried to describe the structure and morphology of our $z\ge6$ 
galaxies using nonparametric methods, including the $CAS$ system, the Gini and 
$M_{20}$ parameters. Compared to low-redshift galaxies, these galaxies occupy 
a smaller range in the parameter space of $CAGM_{20}$. Our simulations show 
that the measurements of these parameters have been systematically biased 
due to the low resolution of the $HST$ images. On the other hand, the relative 
values of these morphological parameters are likely preserved for the vast
majority of galaxies. In addition, we found all expected correlations among 
$CAGM_{20}$ for the $z\ge6$ galaxies. These suggest that the parameters
probably still meaningful for galaxies at similar redshifts, if they are 
measured in the $HST$ images of the same depth and pixel size.

We, for the first time, searched for \lya\ emission halos around $z\ge6$ 
galaxies in narrow-band images. We combined a large number of narrow-band 
images for LAEs at $z\simeq5.7$ and 6.5, respectively. The stacked images 
reached a depth of
$\rm \sim1.2\times10^{-19}\ erg\,s^{-1}\,cm^{-2}\,arcsec^{-2}$ ($1\sigma$).
We did not find evidence of extended diffuse \lya\ emission as predicted 
by cosmological simulations. It is possible that our images are still not 
deep enough to detect \lya\ emission halos, or that the \lya\ halo emission
has been diluted to a much lower level during the construction of the stacked 
images. It is also possible that the halo emission is not as strong as 
predicted. A much larger LAE sample is needed to solve this question.
We also investigated positional differences between the rest-frame UV
continuum emission and \lya\ emission in LAEs, using the $HST$ images and
optical narrow-band images. While in compact LAEs the two positions are well 
consistent, in some merging galaxies show significant positional 
differences, with no preferred directions of offsets. It was explained by
the distribution of the disturbed ISM.

\acknowledgments

Support for this work was provided by NASA through Hubble Fellowship grant 
HST-HF-51291.01 awarded by the Space Telescope Science Institute (STScI), 
which is operated by the Association of Universities for Research in
Astronomy, Inc., for NASA, under contract NAS 5-26555.
L.J., E.E., M.M, and S.C. also acknowledge the support from NASA through 
awards issued by STScI ($HST$ PID: 11149,12329,12616) and by JPL/Caltech 
($Spitzer$ PID: 40026,70094).

{\it Facilities:}
\facility{$HST$ (NICMOS,WFC3)},
\facility{$Spitzer$ (IRAC)},
\facility{$Subaru$ (Suprime-Cam)}

\begin{figure}
\epsscale{0.8}
\plotone{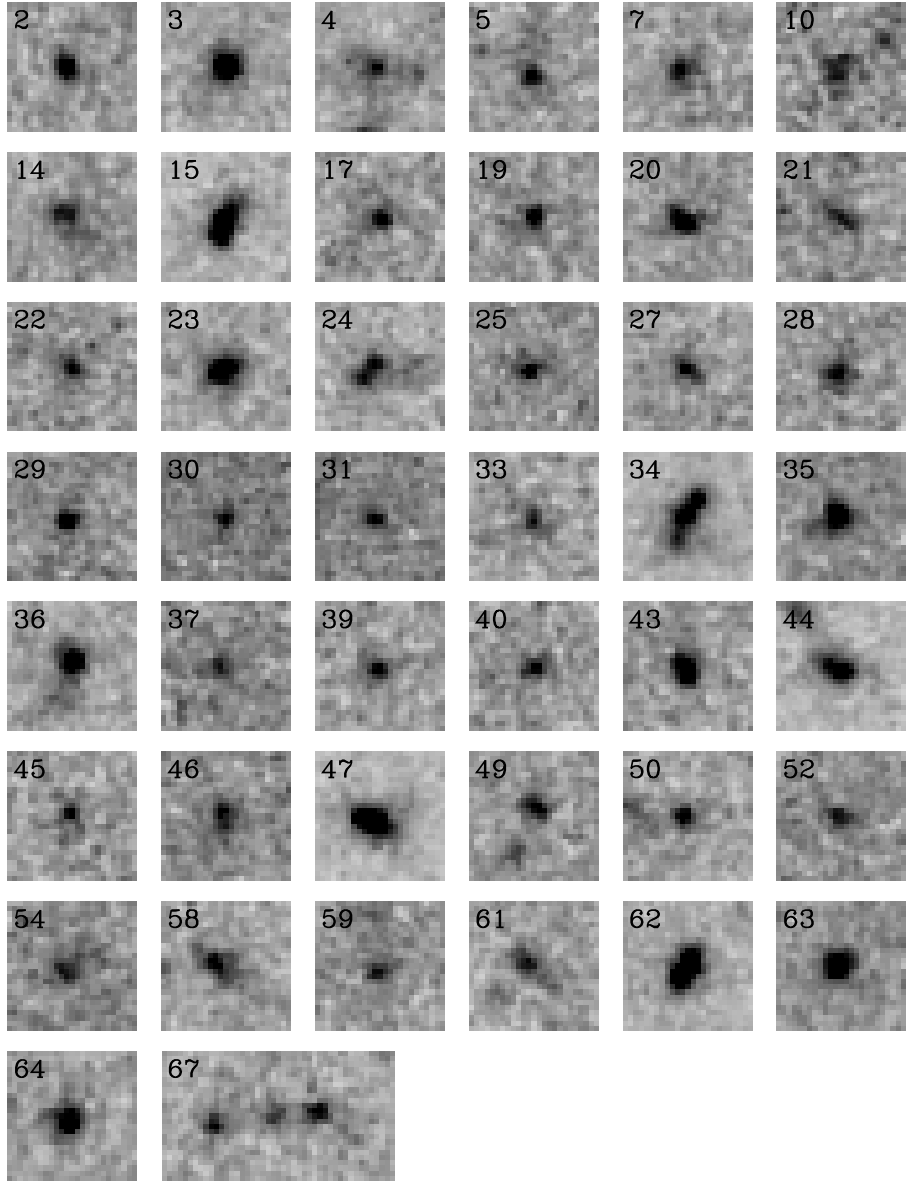}  
\caption{Thumbnail images of the 44 (out of 67) galaxies that have more than 
$10\sigma$ detections in the stacked $HST$ images, shown in order of 
increasing redshift. The numbers of the galaxies plotted in this figure 
correspond to the galaxy numbers given in Table 1 of Paper I. 
The size of the first 43 images is 25 pixel square, which corresponds to 
$2\farcs5 \times 2\farcs5$ in NICMOS images (no. 25, 30, and 31), or 
$1\farcs5 \times 1\farcs5$ in WFC3 images (the rest). The size of no. 67 is 
45 by 25 pixels (or $2\farcs7 \times 1\farcs5$).}
\end{figure}

\clearpage
\begin{figure}
\epsscale{0.6}
\plotone{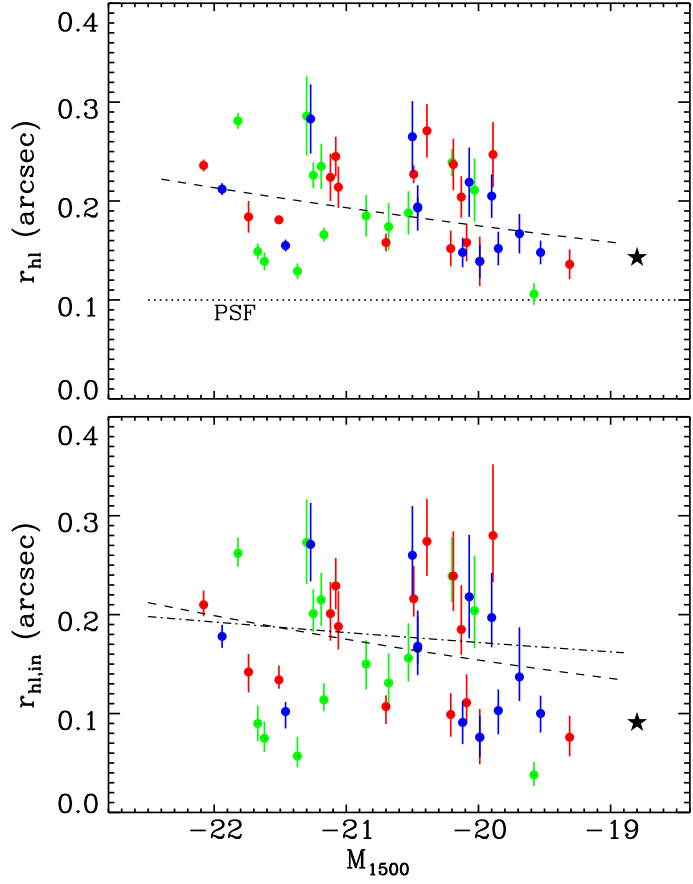}	
\caption{Half-light radius as a function UV continuum luminosity $M_{1500}$.
The blue and red circles represent the LAEs at $z\simeq5.7$ and 6.5 (including
$z\simeq7$), respectively, and the green circles represent the LBGs at
$z\simeq6$. The black star represents the stacked object of the faint 
galaxies, and is arbitrarily put at $M_{1500}=-18.8$ mag.
The upper panel shows the measured radius $r_{\rm hl}$ without correction for 
PSF broadening. The error bars reflect the measurement uncertainties derived
from simulations in Section 3.1. The dotted line indicates the PSF size in our 
$HST$ WFC3 images. The dashed line (best log-linear fit) illustrates the 
weak relation between $r_{\rm hl}$ and $M_{1500}$ 
($r_{\rm hl} \propto L^{0.11 \pm 0.02}$).
The lower panel shows the intrinsic radius $r_{\rm hl,in}$ after correction 
for PSF broadening with simulations in Section 3.1. The error bars include 
both measurement and systematic uncertainties. The dashed line 
(best log-linear fit to all data points) shows the weak relation between 
$r_{\rm hl,in}$ and $M_{1500}$ ($r_{\rm hl,in} \propto L^{0.14 \pm 0.03}$) for
the whole sample. The dash-dotted line (best log-linear fit to the data 
points at $M_{1500}\le -20$ mag) suggests little correlation 
($r_{\rm hl,in} \propto L^{0.06 \pm 0.03}$)
between size and luminosity in luminous galaxies.}
\end{figure}

\clearpage
\begin{figure}
\epsscale{0.6}
\plotone{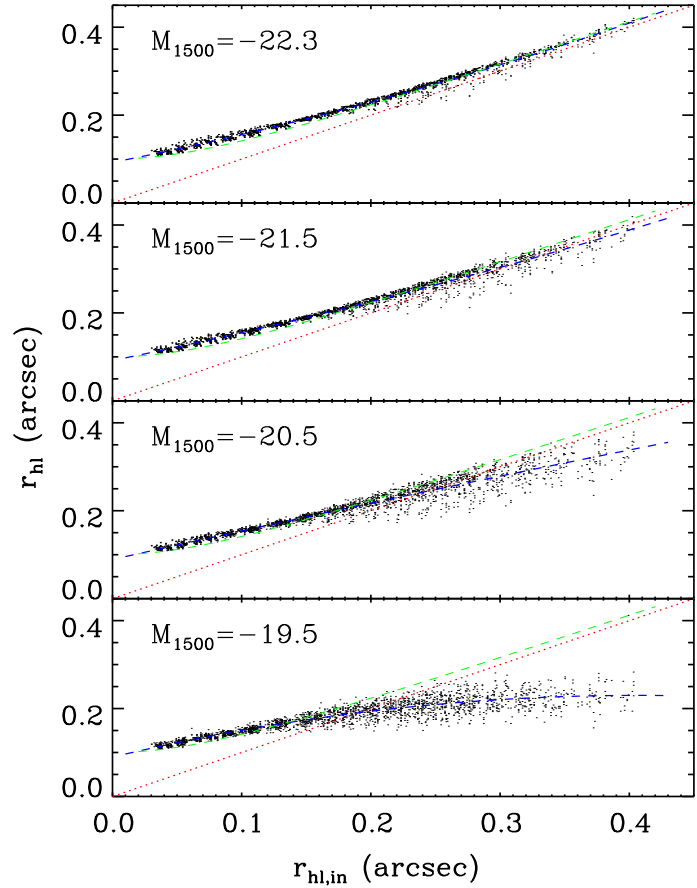}  
\caption{Measured half-light radius $r_{\rm hl}$ as a function of intrinsic
half-light radius $r_{\rm hl,in}$ at four different magnitudes, derived from
our simulations in Section 3.1. The red dotted lines indicate the equality 
relation. At small sizes, $r_{\rm hl}$ is significantly larger than 
$r_{\rm hl,in}$ due to PSF broadening. At large sizes 
($r_{\rm hl,in} \ge 0\farcs2 - 0\farcs3$), $r_{\rm hl}$ starts to fall short 
of $r_{\rm hl,in}$. This happens at smaller sizes for fainter galaxies.
We illustrate this trend by displaying the best second-order polynomial fit to 
the data points (blue dashed lines). The green dashed lines show the relation
of $r^2_{\rm hl,in} = r^2_{\rm hl} - r^2_{\rm PSF}$, which is a good
approximation at small sizes and/or high luminosities, but underestimates
$r_{\rm hl,in}$ elsewhere.}
\end{figure}

\clearpage
\begin{figure}
\epsscale{0.6}
\plotone{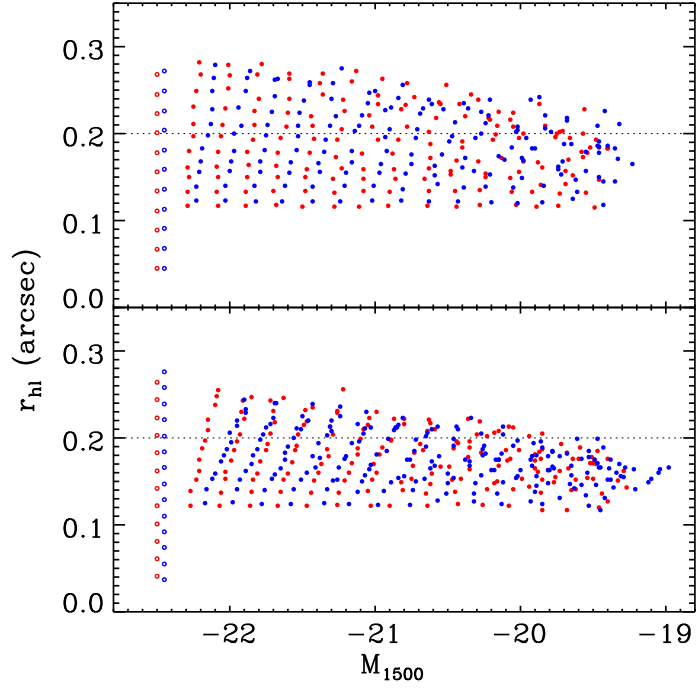}  
\caption{Measured half-light radius $r_{\rm hl}$ as a function of $M_{1500}$ 
for simulated galaxies. The open circles indicate the input intrinsic sizes 
$r_{\rm hl,in}$ of the galaxies in our simulations. The filled circles are the 
measured sizes $r_{\rm hl}$ at different luminosities. The two colors red and 
blue indicate two different $HST$ images that the mock galaxies are placed in. 
The dotted lines are used to guide the eye. The two panels are for two sets of 
S\'{e}rsic index $n$: 1 and 2 times the measured $n$ from our $z\ge6$ 
galaxies. They show that fainter galaxies (with the same intrinsic size) appear 
to be smaller, as also shown in Figure 3. For the same reason, larger galaxies 
(with the same intrinsic luminosity) appear to be slightly fainter
(see discussion in Section 3.1).}
\end{figure}

\clearpage
\begin{figure}
\epsscale{0.8}
\plotone{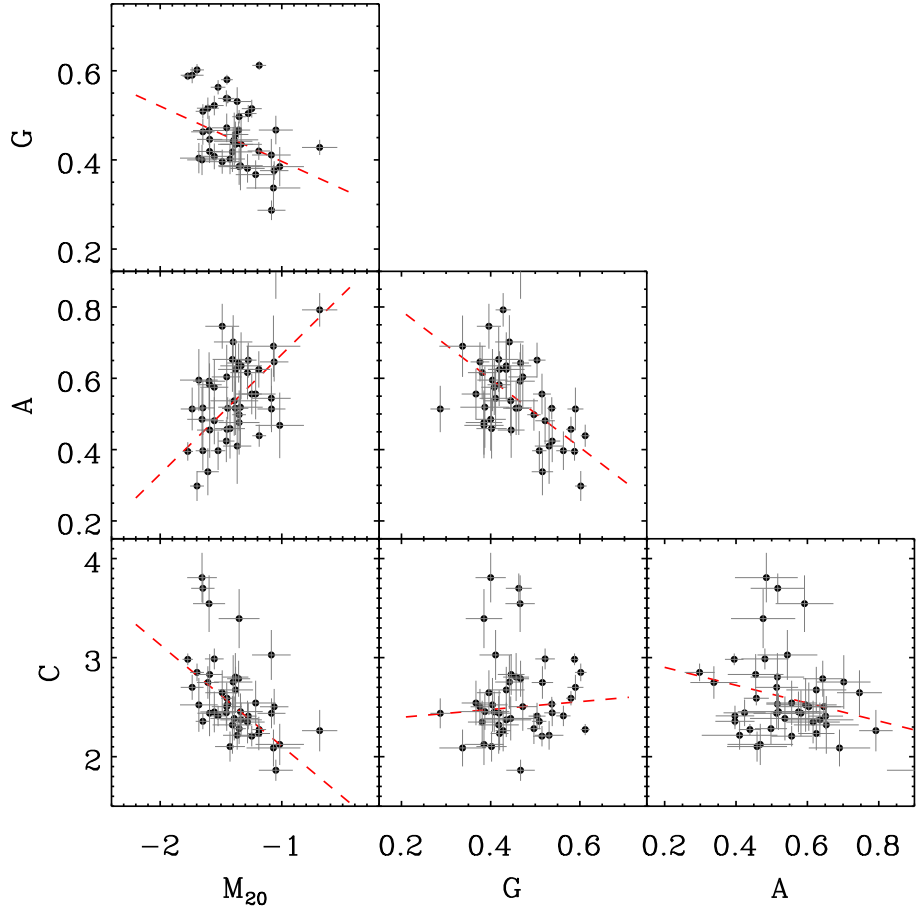}  
\caption{Morphological parameters $CAGM_{20}$ for the galaxies in our sample. 
The measurement uncertainties were estimated from simulations in Section 3.2.1.
Compared to low-redshift galaxies, our galaxies occupy a much narrower range
in the parameter space due to the low resolution of the images.
Nevertheless, each parameter is correlated with one or more of the other 
parameters. The red dashed lines are the best linear fits to the relations.}
\end{figure}

\clearpage
\begin{figure}
\epsscale{0.8}
\plotone{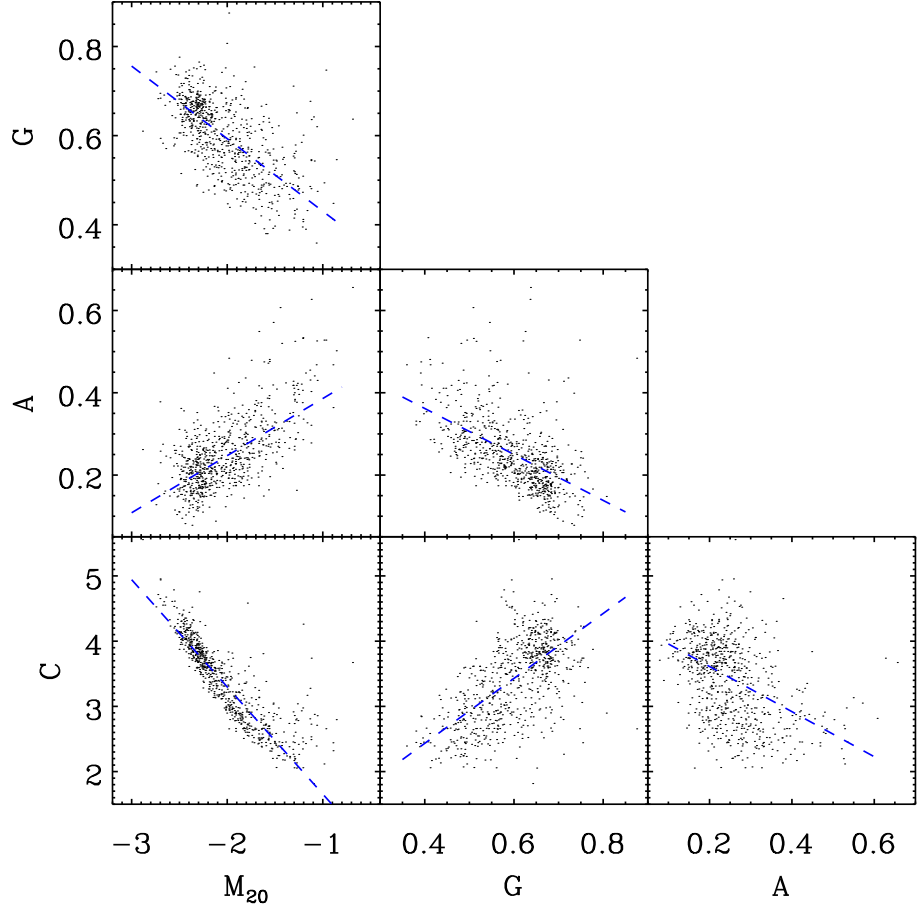}  
\caption{Morphological parameters $CAGM_{20}$ for 740 low-redshift bright 
galaxies selected from the library of galSVM \citep{hue08,hue11}. These 
galaxies cover a large range of the parameter space. The blue dashed lines
are the best linear fits to the data points, showing the correlations among
these parameters. Note that the scale in this figure is very different from
that in Figure 5.}
\end{figure}

\clearpage
\begin{figure}
\epsscale{0.8}
\plotone{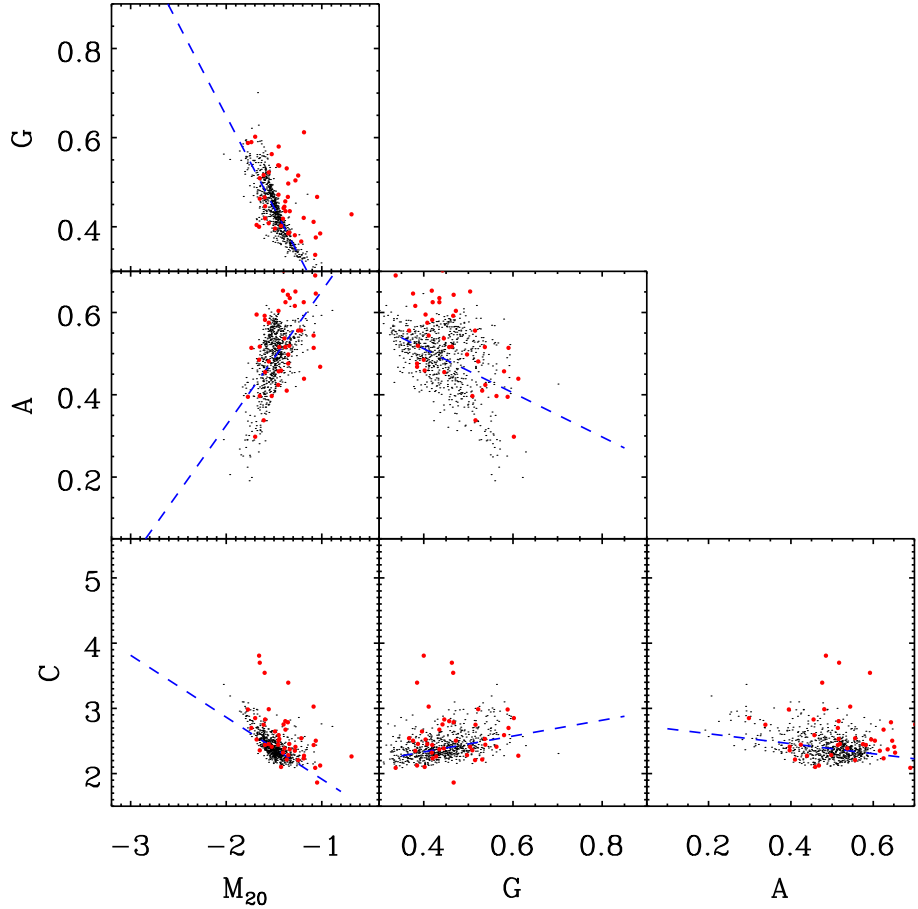}  
\caption{Morphological parameters $CAGM_{20}$ of 740 mock galaxies at $z\ge6$
(black dots) measured in our $HST$ images. The red circles represent our 
$z\ge6$ galaxies (same as Figure 5). The scale in this figure is 
the same as the scale in Figure 6. The mock galaxies were produced 
from the 740 low-redshift galaxies (Figure 6) using simulations in Section 
3.2.2. Compared to Figure 6, the mock galaxies occupy a smaller range of the 
parameter space due to the low resolution of our $HST$ images. 
The blue dashed lines are the best linear fits to the black dots, showing the 
correlations among these parameters. The positions in the parameter space are 
consistent between the $z\ge6$ and mock galaxies, suggesting that the $z\ge6$ 
galaxies are probably not intrinsically less concentrated, nor more 
asymmetric.}
\end{figure}

\clearpage
\begin{figure}
\epsscale{0.6}
\plotone{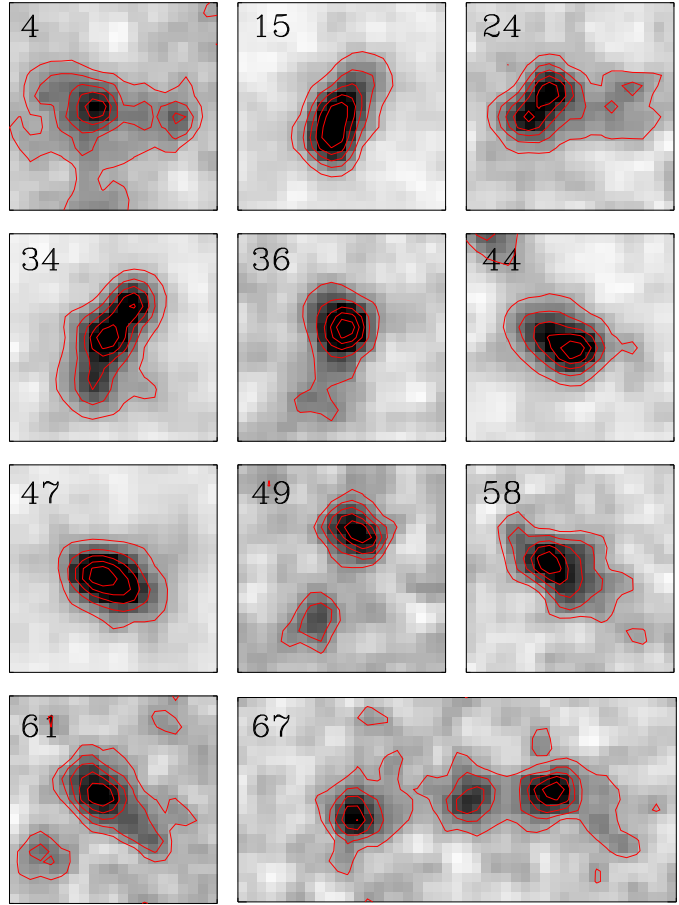}  
\caption{Interacting/merging galaxies identified in our $HST$ images. The 
image size is $1\farcs26 \times 1\farcs26$. The red profiles are the contours 
of SB. Each contour starts at 85\% of the peak value with an interval of 20\%. 
These galaxies have extended and elongated features, and/or have double 
or multiple components.}
\end{figure}

\begin{figure}
\epsscale{0.6}
\plotone{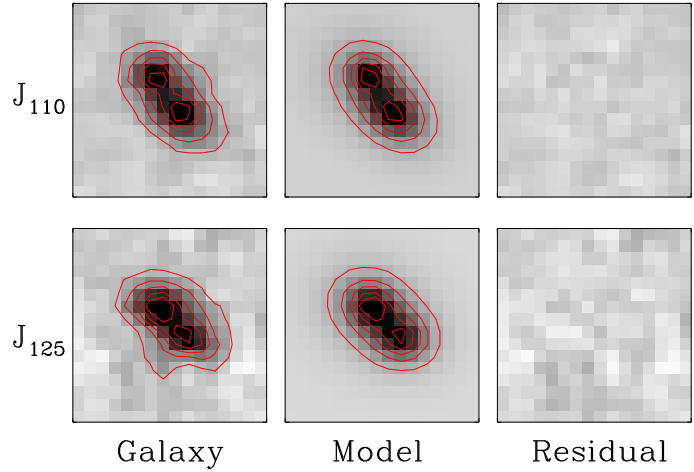}  
\caption{Images of the $z=6.96$ LAE (no. 62). The two left-most images show 
the galaxy in the two $J$ bands ($J_{110}$ and $J_{125}$). 
The two images in the middle are the best-fit model
(two S\'{e}rsic functions for the two components) galaxies. The residuals are
on the right-hand side. The separation between the two components is about
$0\farcs2$ ($\sim1$ kpc).}
\end{figure}

\clearpage
\begin{figure}
\epsscale{0.8}
\plotone{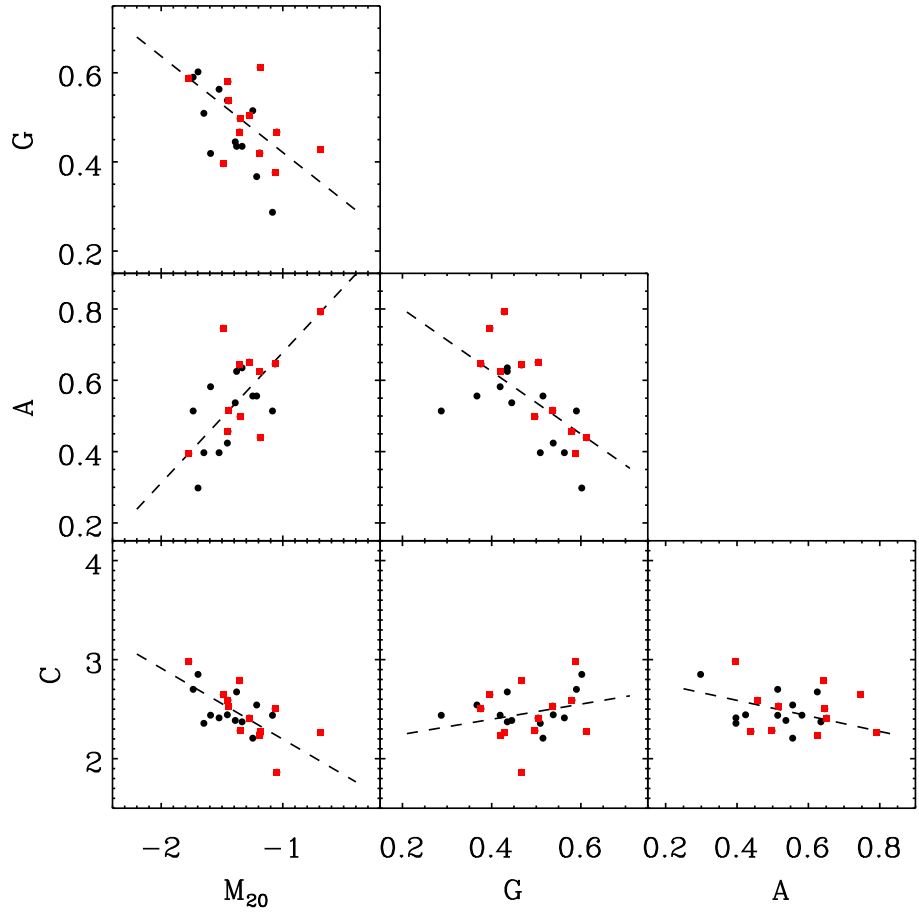}  
\caption{Same as Figure 5, but for galaxies with $M_{1500}\le -20.5$ mag.
The red squares represent the interacting galaxies identified in our sample.
The figure also shows better corrections (smaller scatter) among the 
parameters compared to Figure 5.}
\end{figure}

\clearpage
\begin{figure}
\epsscale{0.55}
\plotone{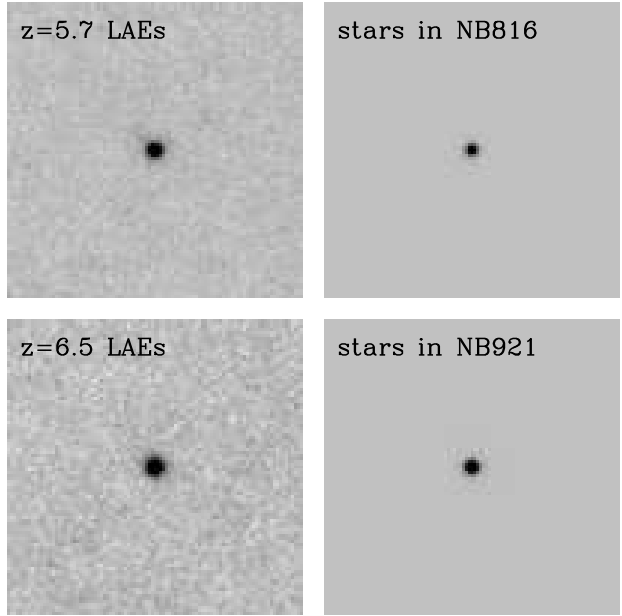}  
\caption{Stacked narrow-band images for LAEs and stars (or point sources) in
two bands NB816 and NB921. The image size is $20\arcsec \times 20\arcsec$.
The images have the same intensity scale. The PSF FWHM sizes derived from the 
two stacked stars are $0\farcs49$ and $0\farcs61$, respectively. The FWHM of
the stacked LAEs are $0\farcs61$ and $0\farcs77$, respectively. They are 
larger than the PSF sizes by 26\%, because LAEs are not point sources.
The stacked LAEs do not show diffuse \lya\ halos.}
\end{figure}

\begin{figure}
\epsscale{0.55}
\plotone{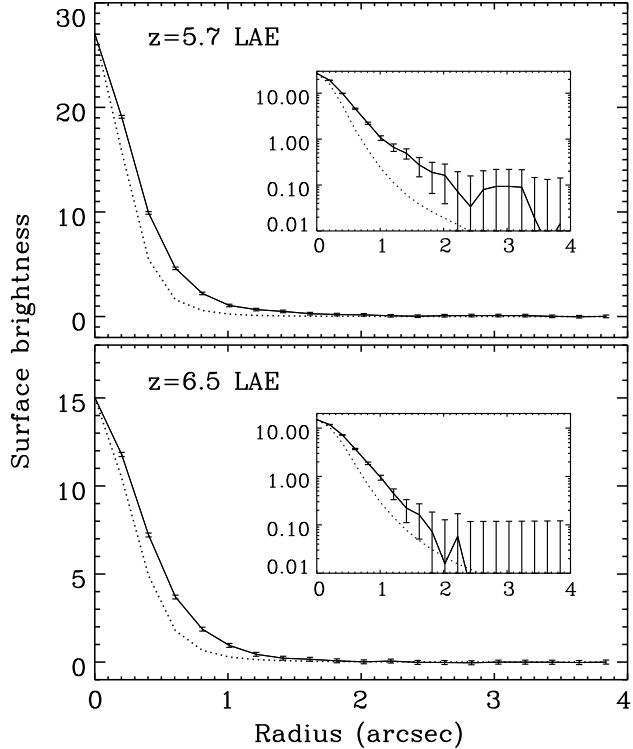}  
\caption{Radial profiles of the stacked images in Figure 11. The units of SB is 
$\rm 10^{-19}\ erg\,s^{-1}\,cm^{-2}\,arcsec^{-2}$. The solid profiles with 
$1\sigma$ error bars represent the stacked LAEs, and the dashed profiles 
represent the stacked stars. The insets show the radial profiles on a log 
scale. The LAE profiles are broader than the PSF sizes, and exhibit slightly 
longer tails than the PSF profiles do, meaning that the \lya\ emission is
resolved. This is because galaxies are not point 
sources, as explained in Section 4.1. At $r\ge2\arcsec$, the SB in both cases 
is consistent with zero within $1\sigma$ errors. We do not see very extended 
halos of \lya\ emission.}
\end{figure}

\clearpage
\begin{figure}
\epsscale{0.6}
\plotone{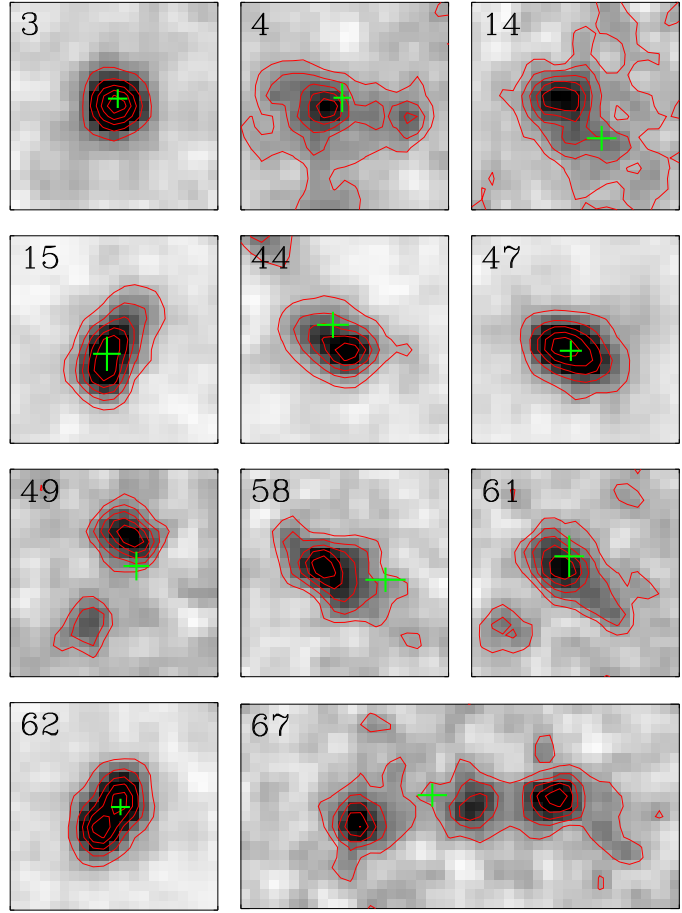}  
\caption{Positions of UV continuum and \lya\ emission for a sample of bright
galaxies. The red contours and their scales are similar to those shown in 
Figure 8. They display the UV continuum emission seen in the $HST$ images. 
The green crosses indicate the positions (and $1\sigma$ uncertainties) of 
the \lya\ emission from our ground-based narrow-band images.
The first object no. 3 represents a typical compact galaxy, whose positions of
UV continuum and \lya\ emission agree with each other. The rest of the objects 
are merging/interacting systems, which show a variety of \lya\ position 
offsets relative to the UV continuum positions, including significant 
positional misalignment.}
\end{figure}


\begin{thebibliography}{}
\bibitem[Abraham et al.(1996)]{abr96} Abraham, R.~G., Tanvir, N.~R., Santiago,
   B.~X., et al.\ 1996, \mnras, 279, L47
\bibitem[Abraham \& van den Bergh(2001)]{abr01} Abraham, R.~G., \&
   van den Bergh, S.\ 2001, Science, 293, 1273
\bibitem[Abraham et al.(2003)]{abr03} Abraham, R.~G., van den Bergh, S., \&
   Nair, P.\ 2003, \apj, 588, 218
\bibitem[Bershady et al.(2000)]{ber00} Bershady, M.~A., Jangren, A., \&
   Conselice, C.~J.\ 2000, \aj, 119, 2645
\bibitem[Bertin \& Arnouts(1996)]{ber96} Bertin, E., \& Arnouts, S.\ 1996,
   \aaps, 117, 393
\bibitem[Blanton \& Moustakas(2009)]{bla09} Blanton, M.~R., \& Moustakas, J.\
   2009, \araa, 47, 159
\bibitem[Bouwens et al.(2006)]{bou06} Bouwens, R.~J., 
	Illingworth, G.~D., Blakeslee, J.~P., \& Franx, M.\ 2006, \apj, 653, 53 
\bibitem[Brinchmann et al.(1998)]{bri98} Brinchmann, J., Abraham, R.,
   Schade, D., et al.\ 1998, \apj, 499, 112
\bibitem[Cai et al.(2011)]{cai11} Cai, Z., Fan, X., Jiang, L., et al.\ 2011,
   \apjl, 736, L28
\bibitem[Carlberg et al.(2000)]{car00} Carlberg, R.~G.,
   Cohen, J.~G., Patton, D.~R., et al.\ 2000, \apjl, 532, L1
\bibitem[Cassata et al.(2010)]{cas10} Cassata, P.,
   Giavalisco, M., Guo, Y., et al.\ 2010, \apjl, 714, L79
\bibitem[Cibinel et al.(2012)]{cib12} Cibinel, A., Carollo, 
	C.~M., Lilly, S.~J., et al.\ 2012, arXiv:1206.6108 
\bibitem[Cole et al.(2000)]{col00} Cole, S., Lacey, C.~G., Baugh, C.~M., \&
   Frenk, C.~S.\ 2000, \mnras, 319, 168
\bibitem[Conselice(2003)]{con03} Conselice, C.~J., 2003, ApJS, 147, 1
\bibitem[Conselice \& Arnold(2009)]{con09} Conselice, C.~J., \& Arnold, J.\
   2009, \mnras, 397, 208
\bibitem[Cooke et al.(2010)]{cooke10} Cooke, J., Berrier, J.~C., Barton, 
	E.~J., Bullock, J.~S., \& Wolfe, A.~M.\ 2010, \mnras, 403, 1020 
\bibitem[Cowie et al.(2011)]{cow11} Cowie, L.~L., Hu, E.~M., \& Songaila,
   A.\ 2011, \apjl, 735, L38
\bibitem[Curtis-Lake et al.(2012)]{cur12} Curtis-Lake, E.,
   McLure, R.~J., Pearce, H.~J., et al.\ 2012, \mnras, 422, 1425
\bibitem[Dijkstra \& Kramer(2012)]{dij12} Dijkstra, M., \& Kramer, R.\ 2012, 
	\mnras, 424, 1672 
\bibitem[Dow-Hygelund et al.(2007)]{dow07} Dow-Hygelund,
   C.~C., Holden, B.~P., Bouwens, R.~J., et al.\ 2007, \apj, 660, 47
\bibitem[Driver et al.(1995)]{driver95} Driver, S.~P.,
	Windhorst, R.~A., Ostrander, E.~J., et al.\ 1995, \apjl, 449, L23 
\bibitem[Driver et al.(1998)]{driver98} Driver, S.~P., 
	Fernandez-Soto, A., Couch, W.~J., et al.\ 1998, \apjl, 496, L93
\bibitem[Driver et al.(2005)]{driver05} Driver, S.~P., Liske, J., 
	Cross, N.~J.~G., De Propris, R., \& Allen, P.~D.\ 2005, \mnras, 360, 81
\bibitem[Fan et al.(2006)]{fan06} Fan, X., Carilli, C.~L., \& Keating, B.\
   2006, \araa, 44, 415
\bibitem[Feldmeier et al.(2013)]{fel13} Feldmeier, J., Hagen, 
	A., Ciardullo, R., et al.\ 2013, arXiv:1301.0462 
\bibitem[Ferguson et al.(2004)]{fer04} Ferguson, H.~C.,
   Dickinson, M., Giavalisco, M., et al.\ 2004, \apjl, 600, L107
\bibitem[Finkelstein et al.(2011)]{fin11} Finkelstein, S.~L., 
	Cohen, S.~H., Windhorst, R.~A., et al.\ 2011, \apj, 735, 5
\bibitem[Furusawa et al.(2008)]{fur08} Furusawa, H., Kosugi, G., Akiyama, M.,
   et al.\ 2008, \apjs, 176, 1
\bibitem[Giavalisco et al.(1996)]{gia96} Giavalisco, M., Steidel, C.~C., \&
   Macchetto, F.~D.\ 1996, \apj, 470, 189
\bibitem[Giavalisco et al.(2004)]{gia04} Giavalisco, M., 
	Ferguson, H.~C., Koekemoer, A.~M., et al.\ 2004, \apjl, 600, L93
\bibitem[Grazian et al.(2012)]{gra12} Grazian, A., Castellano, M., 
	Fontana, A., et al.\ 2012, \aap, 547, A51 
\bibitem[Grogin et al.(2011)]{grogin11} Grogin, N.~A., Kocevski, D.~D., Faber,
   S.~M., et al.\ 2011, \apjs, 197, 35
\bibitem[Gronwall et al.(2011)]{gronwall11} Gronwall, C., Bond,
   N.~A., Ciardullo, R., et al.\ \apj, 743, 9
\bibitem[Hathi et al.(2008)]{hathi08} Hathi, N.~P., Jansen, 
	R.~A., Windhorst, R.~A., et al.\ 2008, \aj, 135, 156 
\bibitem[H{\"a}ussler et al.(2007)]{hau07} H{\"a}ussler, B., 
	McIntosh, D.~H., Barden, M., et al.\ 2007, \apjs, 172, 615
\bibitem[Huertas-Company et al.(2008)]{hue08} Huertas-Company, M., Rouan, D., 
	Tasca, L., Soucail, G., \& Le F{\`e}vre, O.\ 2008, \aap, 478, 971
\bibitem[Huertas-Company et al.(2011)]{hue11} Huertas-Company, M., Aguerri, 
	J.~A.~L., Bernardi, M., Mei, S., \& S{\'a}nchez Almeida, J.\ 2011, \aap, 
	525, A157
\bibitem[Iye et al.(2006)]{iye06} Iye, M., Ota, K.,
   Kashikawa, N., et al.\ 2006, \nat, 443, 186
\bibitem[Jeeson-Daniel et al.(2012)]{jee12} Jeeson-Daniel, 
	A., Ciardi, B., Maio, U., et al.\ 2012, \mnras, 424, 2193 
\bibitem[Jiang et al.(2011)]{jia11} Jiang, L., Egami, E.,
   Kashikawa, N., et al.\ 2011, \apj, 743, 65
\bibitem[Jiang et al.(2013)]{jia13} Jiang, L., Egami, E., Mechtley, M.,
   et al.\ 2013, \apj, in press (arXiv:1303.0024)
\bibitem[Kashikawa et al.(2004)]{kas04} Kashikawa, N.,
   Shimasaku, K., Yasuda, N., et al.\ 2004, \pasj, 56, 1011
\bibitem[Kashikawa et al.(2006)]{kas06} Kashikawa, N.,
   Shimasaku, K., Malkan, M.~A., et al.\ 2006, \apj, 648, 7
\bibitem[Kashikawa et al.(2011)]{kas11} Kashikawa, N., Shimasaku, K.,
   Matsuda, Y., et al.\ 2011, \apj, 734, 119
\bibitem[Kashikawa et al.(2012)]{kas12} Kashikawa, N., Nagao, 
	T., Toshikawa, J., et al.\ 2012, \apj, 761, 85
\bibitem[Koekemoer et al.(2011)]{koe11} Koekemoer, A.~M., Faber, S.~M.,
   Ferguson, H.~C., et al.\ 2011, \apjs, 197, 36
\bibitem[Komatsu et al.(2011)]{kom11} Komatsu, E., Smith,
   K.~M., Dunkley, J., et al.\ 2011, \apjs, 192, 18
\bibitem[Law et al.(2007)]{law07} Law, D.~R., Steidel, C.~C., Erb, D.~K.,
   et al.\ 2007, \apj, 656, 1
\bibitem[Law et al.(2012)]{law12} Law, D.~R., Steidel, C.~C., 
	Shapley, A.~E., et al.\ 2012, \apj, 745, 85 
\bibitem[Le F{\`e}vre et al.(2000)]{lef00} Le F{\`e}vre, O.,
   Abraham, R., Lilly, S.~J., et al.\ 2000, \mnras, 311, 565
\bibitem[Lilly et al.(1998)]{lil98} Lilly, S., Schade, D.,
   Ellis, R., et al.\ 1998, \apj, 500, 75
\bibitem[Lisker(2008)]{lisker08} Lisker, T.\ 2008, \apjs, 179, 319 
\bibitem[Lotz et al.(2004)]{lot04} Lotz, J.~M., Primack, J., \& Madau, P.\
   2004, \aj, 128, 163
\bibitem[Lotz et al.(2006)]{lot06} Lotz, J.~M., Madau, P., Giavalisco, M.,
   Primack, J., \& Ferguson, H.~C.\ 2006, \apj, 636, 592
\bibitem[Lotz et al.(2008)]{lot08} Lotz, J.~M., Davis, M.,
   Faber, S.~M., et al.\ 2008, \apj, 672, 177
\bibitem[Lowenthal et al.(1997)]{low97} Lowenthal, J.~D.,
   Koo, D.~C., Guzman, R., et al.\ 1997, \apj, 481, 673
\bibitem[Malhotra et al.(2012)]{mal12} Malhotra, S., Rhoads, J.~E., 
	Finkelstein, S.~L., et al.\ 2012, \apjl, 750, L36 
\bibitem[Matsuda et al.(2012)]{mat12} Matsuda, Y., Yamada, 
	T., Hayashino, T., et al.\ 2012, \mnras, 425, 878 
\bibitem[Mesinger(2010)]{mes10} Mesinger, A.\ 2010, \mnras, 407, 1328 
\bibitem[Mosleh et al.(2012)]{mos12} Mosleh, M., Williams, R.~J., 
	Franx, M., et al.\ 2012, \apjl, 756, L12 
\bibitem[Nagao et al.(2004)]{nag04} Nagao, T., Taniguchi, Y.,
   Kashikawa, N., et al.\ 2004, \apjl, 613, L9
\bibitem[Nagao et al.(2005)]{nag05} Nagao, T., Kashikawa, N.,
   Malkan, M.~A., et al.\ 2005, \apj, 634, 142
\bibitem[Nagao et al.(2007)]{nag07} Nagao, T.,
   Murayama, T., Maiolino, R., et al.\ 2007, \aap, 468, 877
\bibitem[Oesch et al.(2010)]{oes10} Oesch, P.~A., Bouwens,
   R.~J., Carollo, C.~M., et al.\ 2010, \apjl, 709, L21
\bibitem[Oke \& Gunn(1983)]{oke83} Oke, J. B., \& Gunn, J. E. 1983, \apj,
   266, 713
\bibitem[Ono et al.(2012)]{ono12} Ono, Y., Ouchi, M., 
	Curtis-Lake, E., et al.\ 2012, arXiv:1212.3869 
\bibitem[Ouchi et al.(2008)]{ouc08} Ouchi, M., et al.\ 2008, \apjs, 176, 301
\bibitem[Ouchi et al.(2009)]{ouc09} Ouchi, M., Mobasher, B.,
   Shimasaku, K., et al.\ 2009, \apj, 706, 1136
\bibitem[Ouchi et al.(2010)]{ouc10} Ouchi, M., Shimasaku, K.,
   Furusawa, H., et al.\ 2010, \apj, 723, 869
\bibitem[Ota et al.(2008)]{ota08} Ota, K., Kashikawa, N.,
   Malkan, M.~A., et al.\ 2008, arXiv:0804.3448
\bibitem[Peng et al.(2002)]{pen02} Peng, C.~Y., Ho, L.~C., Impey, C.~D., \&
   Rix, H.-W.\ 2002, \aj, 124, 266
\bibitem[Pirzkal et al.(2007)]{pir07} Pirzkal, N., Malhotra, S.,
   Rhoads, J.~E., \& Xu, C.\ 2007, \apj, 667, 49
\bibitem[Rauch et al.(2011)]{rauch11} Rauch, M., Becker, G.~D.,
	Haehnelt, M.~G., et al.\ 2011, \mnras, 418, 1115
\bibitem[Ravindranath et al.(2006)]{rav06} Ravindranath, S.,
   Giavalisco, M., Ferguson, H.~C., et al.\ 2006, \apj, 652, 963
\bibitem[Schade et al.(1999)]{sch99} Schade, D., et al.\ 1999, \apj, 525, 31
\bibitem[Shi et al.(2009)]{shi09} Shi, Y., Rieke, G., Lotz, J., \& 
   Perez-Gonzalez, P.~G.\ 2009, \apj, 697, 1764 
\bibitem[Shimasaku et al.(2006)]{shi06} Shimasaku, K.,
   Kashikawa, N., Doi, M., et al.\ 2006, \pasj, 58, 313
\bibitem[Stanway et al.(2004)]{sta04} Stanway, E.~R.,
   Glazebrook, K., Bunker, A.~J., et al.\ 2004, \apjl, 604, L13
\bibitem[Stark et al.(2011)]{stark11} Stark, D.~P., Ellis, R.~S., \& Ouchi, 
	M.\ 2011, \apjl, 728, L2
\bibitem[Steidel et al.(2011)]{ste11} Steidel, C.~C., 
	Bogosavljevi{\'c}, M., Shapley, A.~E., et al.\ 2011, \apj, 736, 160
\bibitem[Taniguchi et al.(2005)]{tan05} Taniguchi, Y., Ajiki,
   M., Nagao, T., et al.\ 2005, \pasj, 57, 165
\bibitem[Taniguchi et al.(2009)]{tan09} Taniguchi, Y.,
   Murayama, T., Scoville, N.~Z., et al.\ 2009, \apj, 701, 915
\bibitem[Taylor-Mager et al.(2007)]{taylor07} Taylor-Mager, V.~A., Conselice, 
	C.~J., Windhorst, R.~A., \& Jansen, R.~A.\ 2007, \apj, 659, 162
\bibitem[Toshikawa et al.(2012)]{tos12} Toshikawa, J.,
   Kashikawa, N., Ota, K., et al.\ 2012, \apj, 750, 137
\bibitem[van den Bergh et al.(2000)]{van00} van den Bergh, S., Cohen, J.~G.,
   Hogg, D.~W., \& Blandford, R.\ 2000, \aj, 120, 2190
\bibitem[van der Wel et al.(2012)]{van12} van der Wel, A., 
	Bell, E.~F., H{\"a}ussler, B., et al.\ 2012, \apjs, 203, 24 
\bibitem[Venemans et al.(2005)]{ven05} Venemans, B.~P., R{\"o}ttgering,
   H.~J.~A., Miley, G.~K., et al.\ 2005, \aap, 431, 793
\bibitem[White \& Rees(1978)]{whi78} White, S.~D.~M., \& Rees, M.~J.\ 1978,
   \mnras, 183, 341
\bibitem[Windhorst et al.(2002)]{win02} Windhorst, R.~A., Taylor, V.~A.,
   Jansen, R.~A., et al.\ 2002, \apjs, 143, 113
\bibitem[Windhorst et al.(2008)]{win08} Windhorst, R.~A., Hathi, N.~P., 
	Cohen, S.~H., et al.\ 2008, Advances in Space Research, 41, 1965
\bibitem[Windhorst et al.(2011)]{win11} Windhorst, R.~A., 
	Cohen, S.~H., Hathi, N.~P., et al.\ 2011, \apjs, 193, 27
\bibitem[Yajima et al.(2012)]{yajima12} Yajima, H., Li, Y., Zhu,
   Q., \& Abel, T.\ 2012, \mnras, 424, 884
\bibitem[Zheng et al.(2011)]{zheng11} Zheng, Z., Cen, R., 
	Weinberg, D., Trac, H., \& Miralda-Escud{\'e}, J.\ 2011, \apj, 739, 62
\end{thebibliography}
\end{document}